\documentclass[10pt]{article}
\usepackage[totalheight = 23cm, totalwidth = 17cm]{geometry}
\usepackage{amssymb,amsmath,amsfonts,amsbsy,graphicx,bm}

\def\mathbi#1{\textbf{\em #1}}

\newcommand{\fnl}{f_{\rm NL}}
\newcommand{\calE}{\mathcal{E}}
\newcommand{\calJ}{\mathcal{J}}
\newcommand{\calO}{\mathcal{O}}
\newcommand{\calR}{\mathcal{R}}
\newcommand{\calS}{\mathcal{S}}

\begin{document}

\begin{titlepage}

\rightline{\footnotesize{APCTP-Pre2014-004}} \vspace{-0.2cm}

\begin{center}

\vskip 1.0 cm

{\LARGE  \bf
Non-linear matter bispectrum in general relativity
}

\vskip 1.0cm

{\large
Sang Gyu Biern$^{a}$, Jinn-Ouk Gong$^{b,c}$ and Donghui Jeong$^{d}$
}

\vskip 0.5cm

{\it
$^{a}$Department of Physics, Seoul National University, Seoul 151-747, Korea
\\
$^{b}$Asia Pacific Center for Theoretical Physics, Pohang 790-784, Korea
\\
$^{c}$Department of Physics, Postech, Pohang 790-784, Korea
\\
$^{d}$Department of Physics and Astronomy, Johns Hopkins University, Baltimore, MD 21218, USA
}

\vskip 1.2cm

\end{center}

\begin{abstract}

 We show that the relativistic effects are negligibly small in the non-linear
 density and velocity bispectra. Although the non-linearities of Einstein
 equation introduce additional non-linear terms to the Newtonian fluid
 equations, the corrections to the bispectrum only show up on super-horizon
 scales. We show this with the next-to-leading order non-linear bispectrum
 for a pressureless fluid in a flat Friedmann-Robertson-Walker background,
 by calculating the density and velocity fields up to fourth order.
 We work in the comoving gauge, where the dynamics is identical to the
 Newtonian up to second order.
 We also discuss the leading order matter bispectrum in various gauges,
 and show yet another relativistic effect near horizon scales that the
 matter bispectrum strongly depends on the gauge choice.

\end{abstract}

\end{titlepage}

\newpage
\setcounter{page}{1}

\section{Introduction}
\label{sec:intro}

Recent advances in cosmology has been greatly spurred by precise cosmological
observations. The accurate measurements of the temperature anisotropies
and polarizations in the cosmic microwave background (CMB) by the Wilkinson
Microwave Anisotropy Probe (WMAP) have opened the era of precision
cosmology~\cite{wmap}, and with the most recent PLANCK data we can
constrain the cosmological parameters with less than $\calO(1)$ percent
error~\cite{planck}. With the planned experiments such as
PIXIE~\cite{PIXIE}, PRISM~\cite{PRISM}, and LiteBIRD~\cite{LiteBIRD} to mention a few,
it is guaranteed that we continue our success in the CMB observations and
that we can constrain the cosmological parameters further and can obtain more
information on the early universe as well.

Large scale structure (LSS) of the universe is yet another powerful
cosmological probe, and its importance has ever been increasing
with galaxy surveys such as SDSS~\cite{Anderson:2013zyy},
WiggleZ~\cite{Blake:2011en} and VIPERS~\cite{delaTorre:2013rpa}.
The LSS observations can provide the measurement of geometrical distances,
growth of structures, and shape of primordial correlation functions.
These lower redshift information combined with the CMB data can
break down the degeneracies among cosmological parameters that yields
better constraints than CMB alone~\cite{planck}.
Furthermore, the full three-dimensional information with a huge redshift
coverage available for the LSS observations naturally yields measurement of
properties of dark energy, neutrino properties as well as physics of
the early universe. A number of future observations such as
HETDEX~\cite{HETDEX}, MS-DESI~\cite{MSDESI}, LSST~\cite{LSST}
and Euclid~\cite{Euclid} are proposed to observe LSS with improved accuracy
in near future.

Provided that unprecedentedly accurate data will be soon available in both
CMB and LSS, our theoretical endeavour should also meet the observational
precision. This introduces, however, a number of interesting and important
questions to be addressed, especially for LSS:
\begin{itemize}

\item Non-linearity: With increasing observational accuracy, we can probe the
signal beyond the two-point correlation function in CMB and LSS.
The higher-order correlation functions are the signature of non-linearities.
Searching for the primordial non-Gaussianity~\cite{nGreviews} is a prime example.
The current best constraint from PLANCK is consistent with that the primordial
fluctuations follow the Gaussian statistics with the local non-linearity parameter
$\fnl = 2.7 \pm 5.8$ at $2\sigma$ confidence level.
Non-linearity is more prominent in LSS: gravitational instability
amplifies the density fluctuations to form non-linear structures such as
galaxies and clusters of galaxies.
As a result, the non-linearities deviates the matter power spectrum from the
linear theory predictions~\cite{nlpowerspectrum,Jeong:2010ag}, and generates
large higher-order correlation functions such as bispectrum and trispectrum.
Accurate modeling of non-linearities is, therefore, the key requirement of
exploiting the LSS data at the accuracy level similar to the CMB.

\item Relevance of general relativity:  Most studies on LSS in the past have
been done in the context of the Newtonian gravity~\cite{Bernardeau:2001qr},
which works fine in the small scale, sub-horizon limit.
In order to achieve robust measurements of dark energy properties,
for example from Baryon Acoustic Oscillations (See~\cite{Weinberg:2012es}
for a recent review), planned future LSS surveys will probe larger and larger
volume, and access the scales comparable to the horizon.
Modeling the LSS observables on those large scales demands that we
work in the fully general relativistic context.
The first question that must be addressed is whether
the purely relativistic effects are large enough to be detected or not.
Furthermore, attempted modifications to general relativity (to explain the
recent cosmic acceleration) mostly show up on such very large scales.
Thus LSS is a perfect playground to test modified theories of gravity.

\item Gauge: As we should resort to general relativity, at least in principle,
to study LSS properly, it is crucial to clarify which `gauge' we are using to
interpret the data from LSS surveys. Different gauges are mathematically
equivalent, but it does not mean that physical clarity is also equally shared.
In particular, in the small scale limit the `density contrasts'
$\delta\equiv T^0{}_0/\overline{T}^0{}_0 -1$ in almost all popular gauges are
equivalent to the Newtonian density contrast~\cite{Hwang:2012bi},
but equivalence does not hold on large enough scales close to the horizon.
Of course, by properly choosing the gauge that we interpret the data, the
gauge ambiguity on large scales disappears to yield the gauge invariant
expression for the observable such as the galaxy power spectrum~\cite{gaugedep}.

\end{itemize}
Bearing these in mind, we are encouraged to go beyond the two-point correlation
function or power spectrum, and study the higher-order correlation functions
arisen from the non-linearity in general relativity.

In this article, we study the next-to-leading order non-linearities in the
matter bispectrum in the comoving gauge. The non-linear matter power spectrum
in the same gauge was computed in~\cite{Jeong:2010ag}. In the comoving gauge,
the physical interpretation of the relativistic variables is transparent and
the set of dynamical equations becomes particularly simple. Furthermore, the
equations governing the dynamics of the density and velocity fields exactly coincide
with the usual Newtonian hydrodynamic equations up to second
order~\cite{Noh:2004bc}.
Therefore, the leading order matter bispectrum, which results from
correlating one second order density contrast to two
linear order ones,
in the full relativistic calculation must be the same as that
of the Newtonian calculation, and the purely relativistic contributions
appear from the third order. To obtain the self-consistent next-to-leading
order non-linearities, we calculate the density contrast to the
fourth-order.
We compute the one-loop matter density and velocity bispectra, and confirm
that the purely relativistic corrections are subdominant on cosmologically
relevant scales.

Going beyond the comoving gauge, we also calculate the leading order
matter bispectrum from various other gauges to demonstrate the wild gauge
dependence of the density and velocity bispectra. As in the case for the
galaxy power spectrum, such a gauge dependence should go away when one
calculate the `observable' quantities in each gauge.

This article is organized as follows. In Section~\ref{sec:setup} we present
the perturbation equations of a pressureless matter in the comoving gauge.
In Section~\ref{sec:bispectrum} we give the fourth order solutions of the
perturbation equations in terms of kernels, and compute the matter bispectrum
including one-loop corrections. In Section~\ref{sec:result} we show the total
bispectrum in particular configurations of interest.
In Section~\ref{sec:gauge} we show gauge dependence of the leading bispectrum
in general relativity for large scale study.
We conclude in Section~\ref{sec:conclusion}.

\section{Setup and equations}
\label{sec:setup}

First we present the setup of the background around which we will introduce
density contrast $\delta$ and the peculiar velocity $\bm v$.
We consider a flat Friedmann-Robertson-Walker universe as a background.
Furthermore, to simplify the analysis, we consider the Einstein-de Sitter universe,
i.e. a flat Friedmann model dominated by a pressureless matter.
This is a good enough approximation of our universe at high redshifts.

We find the Arnowitt-Deser-Misner formulation of 3+1
decomposition~\cite{Arnowitt:1962hi} particularly convenient for tracing
the dynamical degrees of freedom in the system.
The four-dimensional line element is given by
\begin{equation}
ds^2 = -N^2dt^2 + \gamma_{ij} \left( N^idt + dx^i \right) \left( N^jdt + dx^j \right) \, ,
\end{equation}
where $N$, $N^i$ and $\gamma_{ij}$ are, respectively, the lapse, shift and
spatial metric.
We use the Roman indices to indicate the spatial dimensions, which are
raised and lowered by the spatial metric.

We only consider scalar perturbations, because the vector and tensor
contributions may be negligibly small on scales where the relativistic effects
are important. To fix the coordinate system, we choose the comoving gauge,
which is defined by
\begin{equation}
\label{comoving}
T^0{}_i = 0 \, .
\end{equation}
This completely fixes the temporal gauge degree of freedom
even at non-linear order~\cite{Noh:2004bc}.
The spatial gauge degree of freedom can be fixed
by taking only a trace component of perturbation in the spatial metric,
\begin{align}
\label{BG_gamma_ij}
\gamma_{ij} = & a^2(1+2\varphi)\delta_{ij} \, .
\end{align}

The comoving gauge condition (\ref{comoving}) gives rise to a particularly
simple form of the energy-momentum tensor $T_{\mu\nu}$. We can write
$T^\mu{}_\nu$ in the perfect fluid form
\begin{equation}
T^\mu{}_\nu = (\rho+p)u^\mu u_\nu + p\delta^\mu{}_\nu \, ,
\end{equation}
from which we can see that the comoving gauge condition demands $u_i=0$.
Then, for a pressureless matter, the energy and momentum densities and the
spatial energy-momentum tensor which appear in the equations we are to solve
are
\begin{align}
\calE & \equiv N^2T^{00} = \rho \, ,
\\
\calJ_i & \equiv NT^0{}_i = 0 \, ,
\\
\calS_{ij} & \equiv T_{ij} = 0 \, .
\end{align}
This will lead to a great simplification of the equations.

Having the setup, we can now write the dynamical equations.
The relevant equations are~\cite{Bardeen:1980kt}
\begin{align}
R^{(3)} + \frac{2}{3}K^2 - \overline{K}^i{}_j\overline{K}^j{}_i & = 2\calE \, ,
\\
\overline{K}^j{}_{i|j} - \frac{2}{3}K_{|i} & = \calJ_i \, ,
\\
\calE_{,0} - N^i\calE_{,i} & = NK\left( \calE + \frac{\calS}{3} \right) + N\overline{K}^i{}_j\overline{S}^j{}_i + \frac{1}{N}\left( N^2\calJ^i \right)_{|i} \, ,
\\
\calJ_{i,0} - N^j\calJ_{i,j} - N^j{}_{,i}\calJ_j & = NK\calJ_i - \left( \calE\delta^j{}_i + \calS^j{}_i \right) N_{|j} - N\calS^j{}_{i|j} \, ,
\\
K_{,0} - N^iK_{,i} & = -N^{|i}{}_{|i} + N \left( R^{(3)} + K^2 + \frac{1}{2}\calS - \frac{3}{2}\calE \right) \, ,
\end{align}
which are, respectively, the energy and momentum constraints,
energy and momentum conservations, and the trace part of the evolution
equation. Here, $R^{(3)}$ is the 3-curvature scalar constructed from
$\gamma_{ij}$, $K$ is the trace of the extrinsic curvature tensor
$K_{ij}\equiv (N_{i|j} + N_{j|i} - \dot{\gamma}_{ij})/N$,
an overbar denotes the traceless part, and a vertical bar denotes
a covariant derivative with respect to $\gamma_{ij}$.

Now, applying our gauge conditions in the Einstein-de Sitter universe, from
the momentum conservation we can see that $N_{,i} = 0$, i.e. the lapse
function is homogeneous. Further, we can identify the perturbation variables
as $\rho = \rho_0 + \delta\rho(t,{\bm x})$ and $K = 3H-\theta(t,{\bm x})$
with $\theta(t,{\bm x}) = \nabla\cdot{\bm v}(t,{\bm x})/a$, so that their
equations in the comoving gauge exactly coincide with the Newtonian continuity
and Euler equations respectively~\cite{Noh:2004bc}. Then, from the energy and momentum
constraint equations we can write respectively $\varphi$ and $N^i$ in terms
of $\delta \equiv \delta\rho/\rho_0$ and ${\bm v}$.
We arrive at the relativistic version of the continuity and Euler equations,
which are up to fourth order:
\begin{align}
\label{continuity}
 \dot\delta + \frac{1}{a}\nabla\cdot\bm{v}
= &
-\frac{1}{a}\nabla\cdot(\delta\bm{v})\nonumber\\
& - \frac{1}{a} \left[ -2\varphi_1\bm{v} + \nabla \left( \Delta^{-1}X_2 \right) \right]\cdot(\nabla\delta)
\nonumber \\
& - \frac{1}{a} \left\{ -2\varphi_2\bm{v} + \nabla \left( \Delta^{-1}X_3 \right) - 2\varphi_1\left[ -2\varphi_1\bm{v} + \nabla \left( \Delta^{-1}X_2 \right) \right] \right\} \cdot(\nabla\delta) \, ,
\end{align}
\begin{align}
\label{euler}
 &-\frac{1}{a}\nabla\cdot \left( \dot{\bm{v}} + H\bm{v} \right) - \frac{\rho_0}{2}\delta\nonumber\\
= & \frac{1}{a^2} \nabla\cdot \left[ (\bm{v}\cdot\nabla)\bm{v} \right]
\nonumber\\
& + \frac{1}{a^2} \bigg( \Delta\left[ (\bm{v}\cdot\nabla)\Delta^{-1}X_2 \right] - (\bm{v}\cdot\nabla)X_2 - \frac{2}{3}X_2(\nabla\cdot\bm{v}) + \frac{2}{3}\varphi_1(\bm{v}\cdot\nabla)(\nabla\cdot\bm{v}) - 4\nabla\cdot \left\{ \varphi_1 \left[ (\bm{v}\cdot\nabla)\bm{v} - \frac{1}{3}(\nabla\cdot\bm{v})\bm{v} \right] \right\} \bigg)
\nonumber\\
& + \frac{1}{a^2} \bigg( \Delta \left[ (\bm{v}\cdot\nabla)\Delta^{-1}X_3 \right] - (\bm{v}\cdot\nabla)X_3 - \frac{2}{3}X_3(\nabla\cdot\bm{v}) + \frac{2}{3}\varphi_2(\bm{v}\cdot\nabla)(\nabla\cdot\bm{v}) - 4\nabla\cdot \left\{ \varphi_2 \left[ (\bm{v}\cdot\nabla)\bm{v} - \frac{1}{3}(\nabla\cdot\bm{v})\bm{v} \right] \right\}
\nonumber\\
& \hspace{1cm} - 4 \nabla\cdot \left[ \varphi_1 \left\{ \left[\nabla\left(\Delta^{-1}X_2\right)\right]\cdot\nabla - \frac{1}{3}X_2 \right\} \bm{v} + \varphi_1 \left\{ \left(\bm{v}\cdot\nabla\right) - \frac{1}{3}\left(\nabla\cdot\bm{v}\right) \right\} \left[\nabla(\Delta^{-1}X_2)\right] \right]
\nonumber\\
& \hspace{1cm} + \frac{2}{3}\varphi_1 \left[ \nabla(\nabla\cdot\bm{v})\cdot\nabla\left( \Delta^{-1}X_2 \right) + 4(\bm{v}\cdot\nabla)X_2 \right] + 12 \nabla \cdot \left\{ \varphi_1^2 \left[ (\bm{v}\cdot\nabla)\bm{v} - \frac{1}{3}(\nabla\cdot\bm{v})\bm{v} \right] \right\} - 4\varphi_1^2(\bm{v}\cdot\nabla)(\nabla\cdot\bm{v})
\nonumber\\
& \hspace{1cm} + 2 \left(\nabla\varphi_1\right)\cdot\left(\nabla\varphi_1\right)\bm{v}\cdot\bm{v} + \frac{2}{3}\left( \nabla\varphi_1\cdot\bm{v} \right)^2 + \nabla\cdot \left\{ \left[ \nabla \left( \Delta^{-1}X_2 \right)\cdot\nabla \right] \nabla \left( \Delta^{-1}X_2 \right) - X_2 \nabla \left(\Delta^{-1}X_2\right) \right\} + \frac{2}{3}X_2^2 \bigg) \, ,
\end{align}
where $\Delta \equiv \delta^{ij}\partial_i\partial_j$ and $\Delta^{-1}$ are spatial Laplacian and inverse Laplacian operators respectively, and
\begin{align}
-\frac{\Delta}{a^2}\varphi_1 = & \frac{\rho_0}{2}\delta - \frac{H}{a}\nabla\cdot\bm{v} \, ,
\\
-\frac{\Delta}{a^2}\varphi_2 = & \frac{1}{4a^2} \left\{ \nabla\cdot \left[ (\bm{v}\cdot\nabla)\bm{v} \right] - (\bm{v}\cdot\nabla)(\nabla\cdot\bm{v}) - (\nabla\cdot\bm{v})^2 \right\} - \frac{1}{2a^2} \left[ 3\left(\nabla\varphi_1\right)\cdot\left(\nabla\varphi_1\right) + 8\varphi_1\Delta\varphi_1 \right] \, ,
\\
X_2 = & 2\varphi_1\nabla\cdot\bm{v} - (\bm{v}\cdot\nabla)\varphi_1 + \frac{3}{2}\Delta^{-1}\nabla\cdot \left[ \Delta\varphi_1\bm{v} + (\bm{v}\cdot\nabla) \left(\nabla\varphi_1\right) \right] \, ,
\\
X_3 = & 2\varphi_1X_2 + 2\varphi_2(\nabla\cdot\bm{v}) - \left(\nabla\varphi_1\right) \cdot \left[\nabla\left(\Delta^{-1}X_2\right)\right] - (\bm{v}\cdot\nabla)\varphi_2
 - 4\varphi_1^2(\nabla\cdot\bm{v}) + 4\varphi_1(\bm{v}\cdot\nabla)\varphi_1
\nonumber\\
& + \frac{3}{2}\Delta^{-1}\nabla\cdot \left[ \Delta\varphi_1\nabla\left(\Delta^{-1}X_2\right) + \Delta\varphi_2\bm{v} + \nabla\left( \Delta^{-1}X_2 \right) \cdot\nabla \left( \nabla\varphi_1 \right) + (\bm{v}\cdot\nabla)\nabla\varphi_2 \right]
\nonumber\\
& - \frac{3}{2}\Delta^{-1}\nabla \cdot \left\{ \left( \nabla\varphi_1 \right) \cdot \left( \nabla\varphi_1 \right)\bm{v} + 3(\bm{v}\cdot\nabla)\varphi_1\nabla\varphi_1 + 4\varphi_1 \left[ (\bm{v}\cdot\nabla)\nabla\varphi_1 + \Delta\varphi_1\bm{v} \right] \right\} \, .
\end{align}
Note that if $\varphi_1=\varphi_2=0$, we recover the Newtonian continuity and Euler equations as can be read from (\ref{continuity}) and (\ref{euler}), respectively. Thus, relativistic contributions are originated from $\varphi_1$ and $\varphi_2$.

\section{One-loop bispectrum}
\label{sec:bispectrum}

\subsection{Solutions}
\label{sec:sol}

We can find the non-linear solutions of (\ref{continuity}) and (\ref{euler})
perturbatively as follows. First the order linear solutions is the same as
the standard ones for the linear perturbation theory,
\begin{align}
\label{lineardelta}
\delta_1({\bm k},t) = & D(t)\delta_1({\bm k},t_0) \, ,
\\
\label{lineartheta}
\theta_1({\bm k},t) = & -aHD(t)\delta_1({\bm k},t_0) \, ,
\end{align}
where $D(t)$ is the linear growth factor which is normalized to unity at
the present time $t=t_0$, and $f\equiv d\log D/d\log a$ is the logarithmic
derivative of the linear growth factor.
Note that $D(t) = a(t)$ in the Einstein de-Sitter universe that we are
considering here.
With these linear solutions for density and velocity, we perturbatively
expand the full non-linear solutions using momentum dependent symmetric
kernels as
\begin{align}
\label{nldelta}
\delta({\bm k},t) = & \sum_{n=1}^\infty \delta_n = \sum_{n=1}^\infty D^n(t) \int \frac{d^3q_1\cdots d^3q_n}{(2\pi)^{3(n-1)}} \delta^{(3)}({\bm k}-{\bm q}_{12\cdots n}) F_n^{(s)}({\bm q}_1, \cdots {\bm q}_n) \delta_1({\bm q}_1)\cdots\delta_1({\bm q}_n) \, ,
\\
\label{nltheta}
\theta({\bm k},t) = & \sum_{n=1}^\infty \theta_n = -aH\sum_{n=1}^\infty D^n(t) \int \frac{d^3q_1\cdots d^3q_n}{(2\pi)^{3(n-1)}} \delta^{(3)}({\bm k}-{\bm q}_{12\cdots n}) G_n^{(s)}({\bm q}_1, \cdots {\bm q}_n) \delta_1({\bm q}_1)\cdots\delta_1({\bm q}_n) \, ,
\end{align}
where $F_1({\bm k}) = G_1({\bm k}) = 1$ and
${\bm q}_{12\cdots n} \equiv \sum_{i=1}^n{\bm q}_i$.
Note that we only consider the fastest growing mode at each order in
perturbations. With this ansatz, (\ref{continuity}) and (\ref{euler}) become
simply differential equations of $F_n$ and $G_n$.
Because the Newtonian hydrodynamical equations are closed at second order
and the relativistic equations coincide with the Newtonian ones up to second order,
the purely relativistic solutions appears from third order.
Note that, in the comoving gauge, purely relativistic terms explicitly include
the comoving horizon scale $k_H \equiv aH$.

The second order kernels are the same as standard perturbation theory~\cite{Bernardeau:2001qr}, and the third order kernels are presented
in (12) and (13) of~\cite{Jeong:2010ag}.
For completeness, we present the equations and solutions for the fourth
order kernels in Appendix~\ref{app:sol}.

\subsection{Tree bispectrum}

The matter bispectrum is defined as
\begin{equation}
\left\langle \delta({\bm k}_1,t)\delta({\bm k}_2,t)\delta({\bm k}_3,t) \right\rangle \equiv (2\pi)^3 \delta^{(3)}(\mathbi{k}_{123}) B({\bm k}_1,{\bm k}_2,{\bm k}_3,t) \, ,
\end{equation}
and the velocity bispectrum is defined in the same way for
$\theta(\mathbi{k},t)$.
Assuming that the linear density perturbation $\delta_1$ follows the Gaussian
statistics, any higher order correlation functions beyond the linear power spectrum $P_{11}$, defined by
\begin{equation}
\left\langle \delta_1({\bm k}_1,t)\delta_1({\bm k}_2,t) \right\rangle \equiv (2\pi)^3 \delta^{(3)}({\bm k}_1+{\bm k}_2) P_{11}(k_1,t) \, ,
\end{equation}
can be written in terms of $P_{11}$. Note that from (\ref{lineartheta}), we can
see that the linear power spectrum of the velocity perturbation is simply
$P_{11}$ multiplied by $k_H^2 \equiv (aH)^2$.
With Gaussian $\delta_1$, the next-to-leading order bispectrum is given by
\begin{align}
& \left\langle \delta({\bm k}_1)\delta({\bm k}_2)\delta({\bm k}_3) \right\rangle
\nonumber\\
= & \Big[ \left\langle \delta_1({\bm k}_1)\delta_1({\bm k}_2)\delta_2({\bm k}_3) \right\rangle + \text{(2 cyclic)} \Big]
\nonumber\\
& + \left\langle \delta_2({\bm k}_1)\delta_2({\bm k}_2)\delta_2({\bm k}_3) \right\rangle + \Big[ \left\langle \delta_1({\bm k}_1)\delta_1({\bm k}_2)\delta_4({\bm k}_3) \right\rangle + \text{(2 cyclic)} \Big] + \Big[ \left\langle \delta_1({\bm k}_1)\delta_2({\bm k}_2)\delta_3({\bm k}_3) \right\rangle + \text{(5 cyclic)} \Big]
\nonumber\\
\equiv & (2\pi)^3 \delta^{(3)}({\bm k}_{123}) \bigg\{ B^{(0)}({\bm k}_1,{\bm k}_2,{\bm k}_3) + \Big[ B^{(1)}_{222}({\bm k}_1,{\bm k}_2,{\bm k}_3) + B^{(1)}_{114}({\bm k}_1,{\bm k}_2,{\bm k}_3) + B^{(1)}_{123}({\bm k}_1,{\bm k}_2,{\bm k}_3) \Big] \bigg\} \, ,
\end{align}
where we have suppressed the time dependence notation.
The leading bispectrum $B^{(0)}$ does not contain any internal momentum
integration, and is thus usually dubbed as the ``tree-level'' bispectrum.
Meanwhile, the leading corrections $B^{(1)}$ all contain one internal
momentum integration and are frequently called as ``one-loop'' corrections.
In the following, we present matter bispectrum only.
The velocity bispectrum is obtained in essentially the same way by replacing
the kernel $G_i$ and supplying the additional factor $-k_H^3$.

We can straightforwardly compute the tree level bispectrum $B^{(0)}$. We first consider $\left\langle \delta_1({\bm k}_1)\delta_1({\bm k}_2)\delta_2({\bm k}_3) \right\rangle$. This reads
\begin{equation}
\left\langle \delta_1({\bm k}_1)\delta_1({\bm k}_2)\delta_2({\bm k}_3) \right\rangle = \int \frac{d^3q_1d^3q_2}{(2\pi)^3} \delta^{(3)}({\bm k}_3-{\bm q}_{12})F_2^{(s)}({\bm q}_1,{\bm q}_2) \Big\langle \delta_1({\bm k}_1)\delta_1({\bm k}_2)\Big[\delta_1({\bm q}_1)\delta_1({\bm q}_2)\Big] \Big\rangle \, .
\end{equation}
Then, we can immediately find
\begin{equation}
\left\langle \delta_1({\bm k}_1)\delta_1({\bm k}_2)\delta_2({\bm k}_3) \right\rangle = (2\pi)^3 \delta^{(3)}({\bm k}_{123}) 2F_2^{(s)}(-{\bm k}_1,-{\bm k}_2)P_{11}(k_1)P_{11}(k_2) \, ,
\end{equation}
and the tree bispectrum is thus
\begin{equation}
\label{Btree}
B^{(0)}({\bm k}_1,{\bm k}_2,{\bm k}_3) = 2F_2^{(s)}(-{\bm k}_1,-{\bm k}_2)P_{11}(k_1)P_{11}(k_2) + \text{(2 cyclic)} \, .
\end{equation}

\subsection{One-loop bispectrum}

\subsubsection{$B^{(1)}_{222}$}

Next we consider the first one-loop correction term, $B^{(1)}_{222}({\bm k}_1,{\bm k}_2,{\bm k}_3)$. From the full expression
\begin{align}
\left\langle \delta_2({\bm k}_1)\delta_2({\bm k}_2)\delta_2({\bm k}_3) \right\rangle =&  \int \frac{d^3q_1\cdots d^3q_2}{(2\pi)^{3\cdot3}} \delta^{(3)}({\bm k}_1-{\bm q}_{12})\delta^{(3)}({\bm k}_2-{\bm q}_{34})\delta^{(3)}({\bm k}_3-{\bm q}_{56})
\nonumber\\
& \times F_2^{(s)}({\bm q}_1,{\bm q}_2)F_2^{(s)}({\bm q}_3,{\bm q}_4)F_2^{(s)}({\bm q}_5,{\bm q}_6)\Big\langle \Big[\delta_1({\bm q}_1)\delta_1({\bm q}_2)\Big] \Big[\delta_1({\bm q}_3)\delta_1({\bm q}_4)\Big] \Big[\delta_1({\bm q}_5)\delta_1({\bm q}_6)\Big] \Big\rangle \, ,
\end{align}
we can find
\begin{align}
B^{(1)}_{222}({\bm k}_1,{\bm k}_2,{\bm k}_3) =& 8 \int \frac{d^3q}{(2\pi)^3}
F_2^{(s)}({\bm q},{\bm k}_1-{\bm q}) F_2^{(s)}(-{\bm q},{\bm k}_2+{\bm q}) F_2^{(s)}(-{\bm k}_1+{\bm q},-{\bm k}_2-{\bm q})\nonumber\\
&\times P_{11}(q)P_{11}(|{\bm k}_1-{\bm q}|)P_{11}(|{\bm k}_2+{\bm q}|) \, .
\end{align}

\subsubsection{$B^{(1)}_{114}$}

For the next term $B^{(1)}_{114}({\bm k}_1,{\bm k}_2,{\bm k}_3)$, we can proceed in the same manner. Let us consider
\begin{align}
\left\langle \delta_1({\bm k}_1)\delta_1({\bm k}_2)\delta_4({\bm k}_3) \right\rangle = &\int \frac{d^3q_1\cdots d^3q_4}{(2\pi)^{3\cdot3}} \delta^{(3)}({\bm k}_3-{\bm q}_{1234})\nonumber\\
&\times
F_4^{(s)}({\bm q}_1,{\bm q}_2,{\bm q}_3,{\bm q}_4)
\Big\langle \delta_1({\bm k}_1)\delta_1({\bm k}_2)
\Big[\delta_1({\bm q}_1)\delta_1({\bm q}_2)\delta_1({\bm q}_3)\delta_1({\bm q}_4)\Big] \Big\rangle \, .
\end{align}
Then after straightforward calculations we find
\begin{align}
B^{(1)}_{114}({\bm k}_1,{\bm k}_2,{\bm k}_3) = 12 \int \frac{d^3q}{(2\pi)^3}
F_4^{(s)}(-{\bm k}_1,-{\bm k}_2,{\bm q},-{\bm q}) P_{11}(k_1)P_{11}(k_2)P_{11}(q) + \text{(2 cyclic)} \, .
\end{align}

\subsubsection{$B^{(1)}_{123}$}

Finally, we consider the last contribution $B^{(1)}_{123}({\bm k}_1,{\bm k}_2,{\bm k}_3)$ with
\begin{align}\label{B123}
\left\langle \delta_1({\bm k}_1)\delta_2({\bm k}_2)\delta_3({\bm k}_3) \right\rangle = & \int \frac{d^3q_1\cdots d^3q_5}{(2\pi)^{3\cdot3}}
\delta^{(3)}({\bm k}_2-{\bm q}_{12})\delta^{(3)}({\bm k}_3-{\bm q}_{345})
\nonumber\\
& \times
F_2^{(s)}({\bm q}_1,{\bm q}_2)F_3^{(s)}({\bm q}_3,{\bm q}_4,{\bm q}_5)\Big\langle \delta_1({\bm k}_1) \Big[\delta_1({\bm q}_1)\delta_1({\bm q}_2)\Big]
\Big[\delta_1({\bm q}_3)\delta_1({\bm q}_4)\delta_1({\bm q}_5)\Big] \Big\rangle \, .
\end{align}
There are two different ways of correlating the six $\delta_1$'s.
Let us call them ($a$) and ($b$). First, ($a$) is that the two propagators
from $\delta_2$ vertex are connected to both $\delta_1$ and $\delta_3$
vertices, and the remaining two propagators within $\delta_3$ are
inter-connected and form a loop. And ($b$) is that the two propagators
from $\delta_2$ vertex are both connected to $\delta_3$ vertex, and the
remaining one propagator of $\delta_3$ is connected to $\delta_1$ vertex.
That is, in terms of momentum shown in (\ref{B123}), ($a$) corresponds
to the correlations that one of $\{{\bm q}_1,{\bm q}_2\}$ is correlated
to ${\bm k}_1$, and the remaining one is correlated to
$\{{\bm q}_3,{\bm q}_4,{\bm q}_5\}$.
For ($b$), two of \ $\{{\bm q}_3,{\bm q}_4,{\bm q}_5\}$ are correlated
to $\{{\bm q}_1,{\bm q}_2\}$, and the
remaining one is correlated to ${\bm k}_1$.
There are six non-zero contributions for each of ($a$) and ($b$),
and we work out to find
\begin{align}
B^{(1)}_{123}({\bm k}_1,{\bm k}_2,{\bm k}_3) = & B^{(1)}_{123a}({\bm k}_1,{\bm k}_2,{\bm k}_3) + B^{(1)}_{123b}({\bm k}_1,{\bm k}_2,{\bm k}_3) \, ,
\\
B^{(1)}_{123a}({\bm k}_1,{\bm k}_2,{\bm k}_3) = & 6F_2^{(s)}(-{\bm k}_1,-{\bm k}_3) \int \frac{d^3q}{(2\pi)^3} F_3^{(s)}({\bm k}_3,{\bm q},-{\bm q}) P_{11}(k_1)P_{11}(k_3)P_{11}(q) + \text{(5 cyclic)} \, ,
\\
B^{(1)}_{123b}({\bm k}_1,{\bm k}_2,{\bm k}_3) = & 6 \int \frac{d^3q}{(2\pi)^3} F_2^{(s)}({\bm q},{\bm k}_2-{\bm q}) F_3^{(s)}(-{\bm k}_1,-{\bm k}_2+{\bm q},-{\bm q}) P_{11}(k_1)P_{11}(q)P_{11}(|{\bm k}_2-{\bm q}|)
+ \text{(5 cyclic)} \, .
\end{align}
One can find the diagramatic representation of the one-loop bispectrum
in \cite{Scoccimarro:1996jy}.

\section{Results}
\label{sec:result}

To highlight the general relativistic effect at one-loop level, we find it
sufficient to show some special triangular configurations.
We set the three momenta $\mathbi{k}_1$, $\mathbi{k}_2$ and
$\mathbi{k}_3$ in such a way that $|\mathbi{k}_1| = |\mathbi{k}_2| = k$
and $|\mathbi{k}_3| = k/\alpha$, and vary $\alpha$ for different configurations
of interest.
For example, $\alpha = 1/2$, $\alpha = 1$ and $\alpha \gg 1$ correspond to
the folded, equilateral and squeezed configurations, respectively.
We implement the integration in the one-loop calculation by setting
$\mathbi{k}_1$, $\mathbi{k}_2$ and $\mathbi{k}_3$ on the $xz$ plane
with $\mathbi{k}_1$ being aligned along the $z$ axis. To perform the integration over $\mathbi{q}$, we introduce the magnitude of $\mathbi{q}$ and the cosine between $\mathbi{q}$ and $\mathbi{k}_1$ as $q = rk$ and $\mathbi{k}_1\cdot\mathbi{q} = k^2r\mu$ with $0 \leq r \leq \infty$ and $-1 \leq \mu \leq 1$. Then, each vector including the internal momentum $\mathbi{q}$ is given respectively by
\begin{align}
\mathbi{k}_1 & = (0,0,k) \, ,
\\
\mathbi{k}_2 & = \left( \frac{\sqrt{4\alpha^2-1}}{2\alpha^2}k, 0, \frac{1-2\alpha^2}{2\alpha^2}k \right) \, ,
\\
\mathbi{k}_3 & = -\mathbi{k}_1-\mathbi{k}_3 \, ,
\\
\mathbi{q} & = \left( kr\sqrt{1-\mu^2}\cos\phi , kr\sqrt{1-\mu^2}\sin\phi , kr\mu  \right) \, .
\end{align}
We calculate the linear matter power spectrum $P_{11}(k)$ from \texttt{CAMB}
code with cosmological parameters given in Table 1 of~\cite{Komatsu:2010fb}.
We find that setting the radial integral range for $r$ from
$r_\text{min}=10^{-2}k_H/k$ to $r_\text{max}=10^6k_H/k$ is sufficient to guarantee
the convergence. All results we show hereafter are for $z=0$.

\begin{figure}[h]
  \centering
  \includegraphics[width=6.5in]{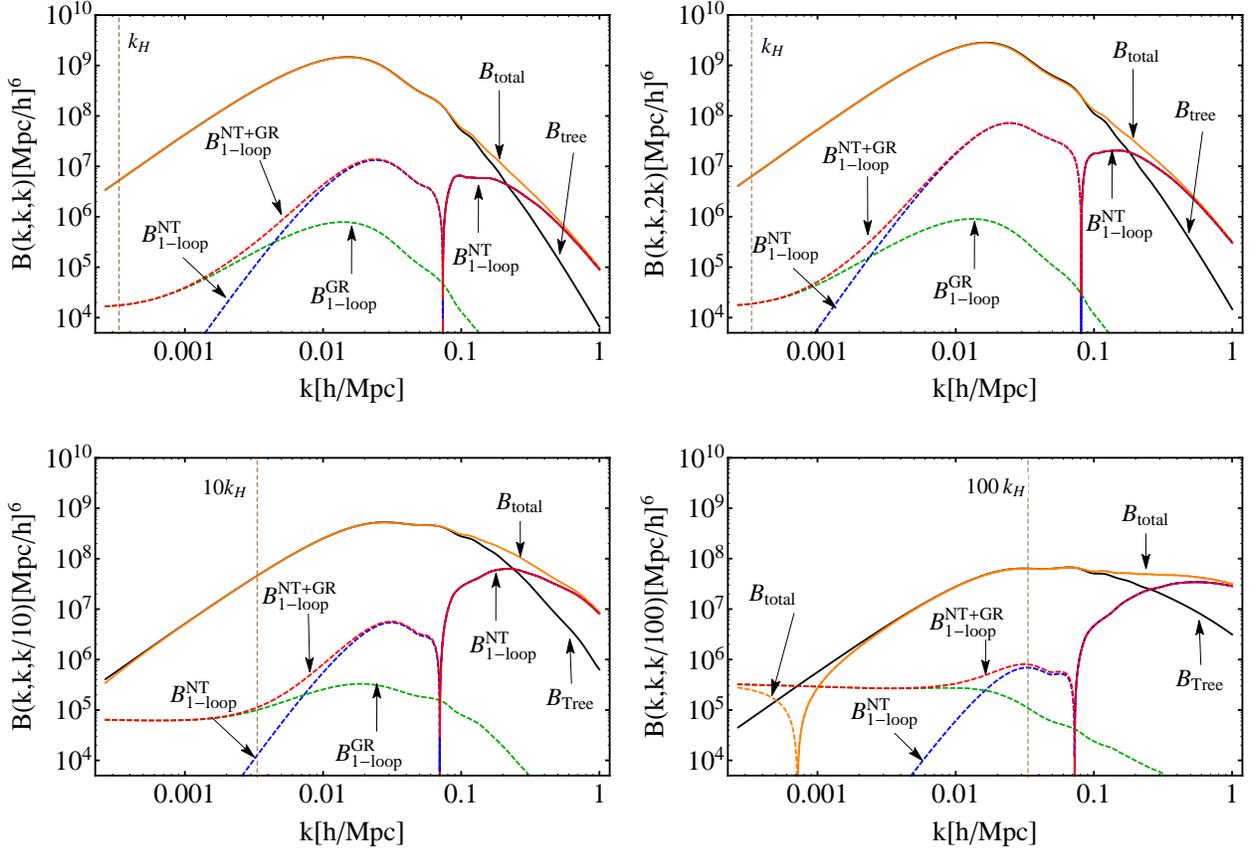}
  \caption{
In each panel, we present the matter density bispectrum in the (clockwise from top left) equilateral, folded, tightly ($\alpha=100$) and slightly ($\alpha=10$) squeezed configuration at $z=0$. Solid (dashed) lines indicate that the corresponding contributions have positive (negative) values. The vertical dotted line denotes the Hubble horizon scale $k_H$.}
  \label{fig:deltabi}
\end{figure}

In Figure~\ref{fig:deltabi} we show the matter bispectrum $B_{\rm total}$
up to one-loop corrections as well as individual component:
leading order $B_{\rm tree}$, Newtonian one-loop $B_{1-{\rm loop}}^{\rm NT}$,
relativistic one-loop $B_{1-{\rm loop}}^{\rm GR}$ and their sum
$B_{1-{\rm loop}}^{\rm NT + GR}$. For each curve, dashes lines show the
absolute value of the negative quantity.
The Newtonian one-loop corrections are appreciable on sub-horizon scales,
$k \gtrsim 0.1h$Mpc$^{-1}$, and dominates the tree contribution for large $k$
indicating the strong non-linearities due to gravitational instability.
They change sign at around $k \sim 0.1h$Mpc$^{-1}$, and on smaller (larger)
scales the Newtonian corrections are negative (positive).
The general relativistic one-loop corrections are strongly suppressed on
small scales, but we note that on very large scales ($k\to0$ limit)
they approach a constant value.
While sub-dominant on all scales in the equilateral and folded
configurations, the relativistic corrections give rise to the notable
changes to the matter bispectrum on large scales for more squeezed triangles.
In the tightly squeezed limit ($\alpha=100$) they even dominate the tree
contribution and make the total bispectrum negative, i.e. anti-correlated.
This peculiar behavior is mainly coming from the components that carry
$k_H^4$ factor in the fourth order kernel $F_4$.

We estimate the behavior of the large scale plateau as the following.
For simplicity, let us abbreviate the radial integration
with the $k_H^4$ factor of $F_4$ in $B_{114}^{(1)}(k,k,k/\alpha)$ as
$\int dr P_{11}(kr)f(r)$. The variable $r \equiv q/k$ is very large on large scale since $k\to0$.
With this, we can understand the asymptotic behavior of $f(r)$ on large scale ($r\gg1$) and
in the squeezed limit ($\alpha\gg1$) as
\begin{equation}
f(r) = -\frac{75k_H^4}{28\pi^2k}\alpha^2 P_{11}(k)P_{11}\left(\frac{k}{\alpha}\right) + \calO(r^{-2})\, .
\end{equation}
Note that $f(r)$ is to leading order independent of $r$ in the large scale limit. Using the fact $P_{11}(k)\propto k^{n_\calR}$ for small $k$ with $n_\calR \sim 0.96$ being the spectral index of the primordial perturbation, we can simply write the squeezed bispectrum on very large scales as
\begin{equation}
B_{114}^{(1)}(k,k,k/\alpha) \propto -\frac{75k_H^4}{28\pi^2}\alpha k^{2(n_\calR-1)} \, .
\end{equation}
Therefore, the matter bispectrum in the squeezed configuration is proportional
to $\alpha$ and is nearly independent of $k$.

\begin{figure}[!h]
  \centering
  \includegraphics[width=6.5in]{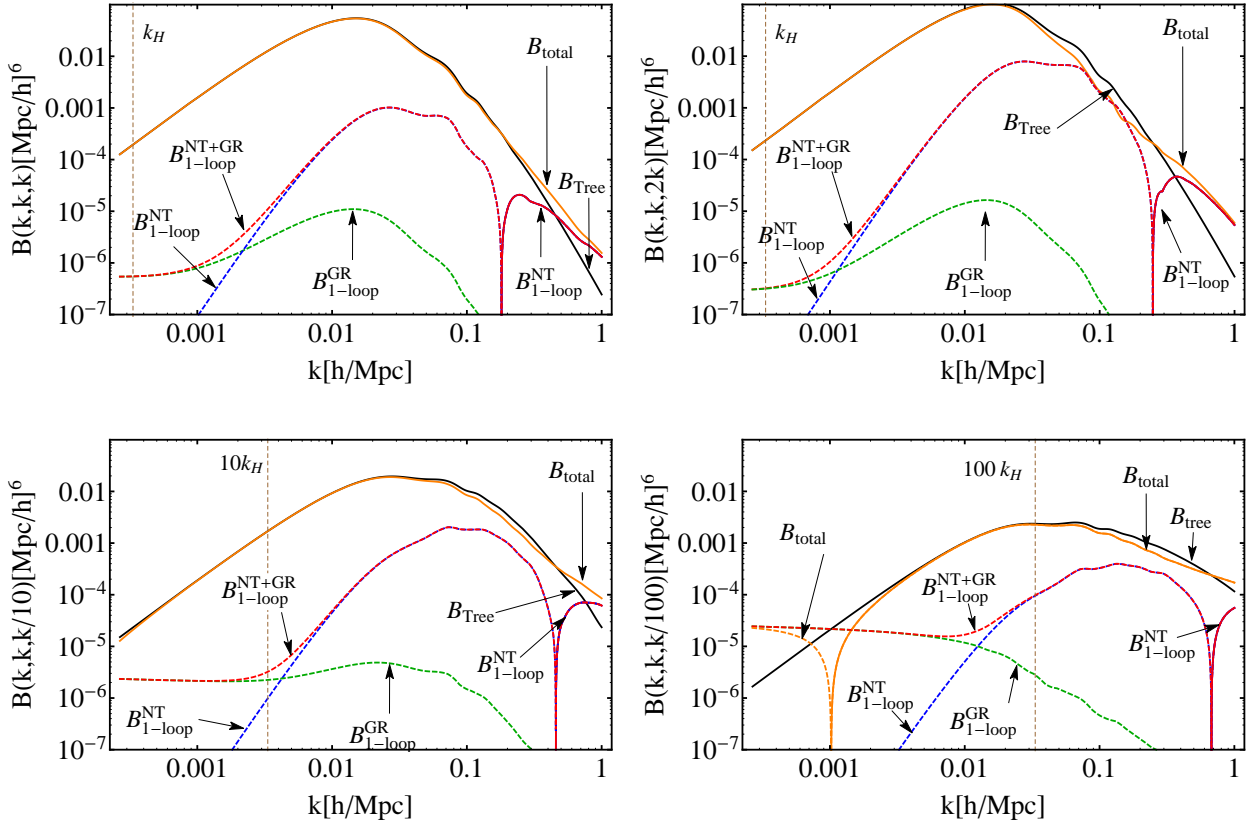}
\caption{
Velocity bispectrum shown in the same manner as Figure~\ref{fig:deltabi}.}
  \label{fig:thetabi}
\end{figure}

We show in Figure~\ref{fig:thetabi} the velocity bispectrum. The non-linear
velocity bispectrum shows the similar features as the density bispectrum in
Figure~\ref{fig:deltabi}.
Especially, the plateau on large scales in the squeezed configuration can be
estimated in a similar way: writing the relevant component from $G_4$
schematically as $k_H^3\int dr P_{11}(kr)g(r)$, in the large scale limit we can find
$g(r) = f(r)/3$.
Note, however, that the magnitude is much smaller than the
matter bispectrum, due to the suppression by a factor of $k_H^3$.

As we see from both figures, the relativistic corrections to the density and
velocity bispectra are very well-regulated, as the general relativistic
signature shows only with a small amplitude on smaller scales and is
noticible only for the large scale where the smallest mode is beyond the
horizon scale, $k_H=aH$, shown as a vertical dashed line in the figures.

\section{Tree bispectrum in other gauges}
\label{sec:gauge}

As we have seen above, the matter and velocity bispectra are well-regulated
on all scales, and the relativistic effects are noticeable only on large scales
beyond the horizon scale.
This is because the comoving gauge is privileged in such a way that we have
the same results as the Newtonian calculation up to second order.
However, in other gauges this is not guaranteed and we in general expect
deviations from each other, especially on large scales. To illustrate
this point, we show in a few popular gauges the tree level bispectrum for which
we need second order perturbation $\delta_2$, or equivalently, second order
kernel $F_2$. 
Note that, while we calculate the second order solutions with various gauge choices, the second order kernels $F_2$ we present here are still defined in Eq.~(\ref{nldelta}) with the linear density contrast in the comoving gauge that we are referring as $\delta_1$ throughout this paper.

\begin{itemize}

 \item Comoving gauge:
 This is the main gauge we work with in this article. As we have worked out
 in the previous section, $\delta_1$ and $\delta_2$ are identical to the
 Newtonian density perturbations in the Eulerian coordinate.
 Thus there is no relativistic contributions at tree level.

\item Synchronous gauge:
 Synchronous gauge takes no perturbation in the $00$ and $0i$ components of the metric,
 \begin{equation}
 ds^2 = -dt^2 + a^2(1+2\varphi)\delta_{ij}dx^idx^j \, ,
 \end{equation}
 so that the time coordinate agrees with proper time. In this gauge $\delta_1$ is the same as that in the comoving gauge, but the second order kernel is found to be
 \begin{equation}
 F_2^\text{(sg)}(\mathbi{q}_1,\mathbi{q}_2) = \frac{5}{7}
+ \frac{2}{7}\frac{(\mathbi{q}_1\cdot\mathbi{q}_2)^2}{q_1^2q_2^2} \, .
 \end{equation}
 Thus although there is no divergence on large scales, the tree bispectrum does not match that in the comoving gauge everywhere.
 This is because the density field in the synchronous gauge can be interpreted
 as the Newtonian density perturbation in the Lagrangian point of view following the moving volume elements.

 \item Zero shear gauge:
 In this gauge the metric is written as
 \begin{equation}
 ds^2 = -(1+2\Phi)dt^2 + a^2(1-2\Psi)\delta_{ij}dx^idx^j \, .
 \end{equation}
 We find the linear perturbation is given by
 \begin{equation}
 \delta^\text{(zsg)}_1(\mathbi{k}) = \left( 1+\frac{3k_H^2}{k^2} \right) \delta_1(\mathbi{k}) \, .
 \end{equation}
 Thus, while we recover the same result on sub-horizon scales as in the comoving gauge, deviation becomes prominent as we approach the horizon scale and eventually we face divergence on super-horizon scales. We can find the second order kernel as
 \begin{align}
  F_2&^\text{(zsg)}(\mathbi{q}_1,\mathbi{q}_2) =  F_2^\text{(cg)}(\mathbi{q}_1,\mathbi{q}_2)
  \nonumber\\
  & + \left( \frac{k_H}{k} \right)^2 \left[ \frac{9}{7} + \frac{q_{12}^2}{q_1^2} + \frac{q_{12}^2}{q_2^2} + \frac{12 \left(\mathbi{q}_1\cdot\mathbi{q}_2\right)^2}{7 q_1^2 q_2^2} + \frac{3\mathbi{q}_1\cdot\mathbi{q}_2}{2 q_1^2} + \frac{3\mathbi{q}_1\cdot\mathbi{q}_2}{2 q_2^2} \right]
  \nonumber\\
  & + \left(\frac{k_H}{k}\right)^4 \left[ \frac{105}{4} - \frac{15 \left(\mathbi{q}_1\cdot\mathbi{q}_2\right)^2}{4 q_1^2 q_2^2} + \frac{75 \mathbi{q}_1\cdot\mathbi{q}_2}{4 q_1^2} + \frac{75 \mathbi{q}_1\cdot\mathbi{q}_2}{4 q_2^2} + \frac{15 q_2^2}{2 q_1^2} + \frac{15 q_1^2}{2 q_2^2} + \frac{9q_{12}^4}{2q_1^2q_2^2} + \frac{3q_{12}^2\mathbi{q}_1\cdot\mathbi{q}_2}{2q_1^2q_2^2} \right] \, .
  \end{align}
  This also matches the comoving gauge kernel on small scales but diverges
in the limit $k\to0$. The relativistic effect of gauge dependence in this gauge
is characterized by the second and the third terms in the kernel with a
factor of $k_H$.

 \item Uniform curvature gauge: This gauge is also called as the flat gauge. In this gauge the spatial metric is set to be unperturbed,
 \begin{equation}
 ds^2 = -(1+2A)dt^2 - 2B_i dx^i dt + a^2\delta_{ij}dx^idx^j \, .
 \end{equation}
 The linear perturbation in this gauge is
 \begin{equation}
 \delta^\text{(ucg)}_1(\mathbi{k}) = \left( 1+\frac{15k_H^2}{2k^2} \right) \delta_1(\mathbi{k}) \, ,
 \end{equation}
 thus as in the zero shear gauge we find divergence on large scales. The second order kernel is given by
 \begin{align}
   F_2^\text{(ucg)}(\mathbi{q}_1,\mathbi{q}_2) = & F_2^\text{(cg)}(\mathbi{q}_1,\mathbi{q}_2)
   \nonumber\\
   & + \left( \frac{k_H}{k} \right)^2 \left[ \frac{15}{4} + \frac{5q_{12}^2}{2q_1^2} + \frac{5q_{12}^2}{2q_2^2} + \frac{15 \left(\mathbi{q}_1\cdot\mathbi{q}_2\right)^2}{4 q_1^2 q_2^2} + \frac{15 \mathbi{q}_1\cdot\mathbi{q}_2}{4 q_1^2}+\frac{15 \mathbi{q}_1\cdot\mathbi{q}_2}{4 q_2^2} \right]
   \nonumber\\
   & + \left( \frac{k_H}{k} \right)^4\left(\frac{225q_{12}^2 \mathbi{q}_1\cdot\mathbi{q}_2}{8 q_1^2 q_2^2}+\frac{75q_{12}^2}{2 q_1^2}+\frac{75q_{12}^2}{2 q_2^2}+\frac{75q_{12}^4}{8q_1^2q_2^2}\right) \, .
  \end{align}
  Likewise, we recover the comoving gauge result on small scales $k\to\infty$.

\end{itemize}

\begin{figure}[!h]
\centering
\includegraphics[width=6.5in]{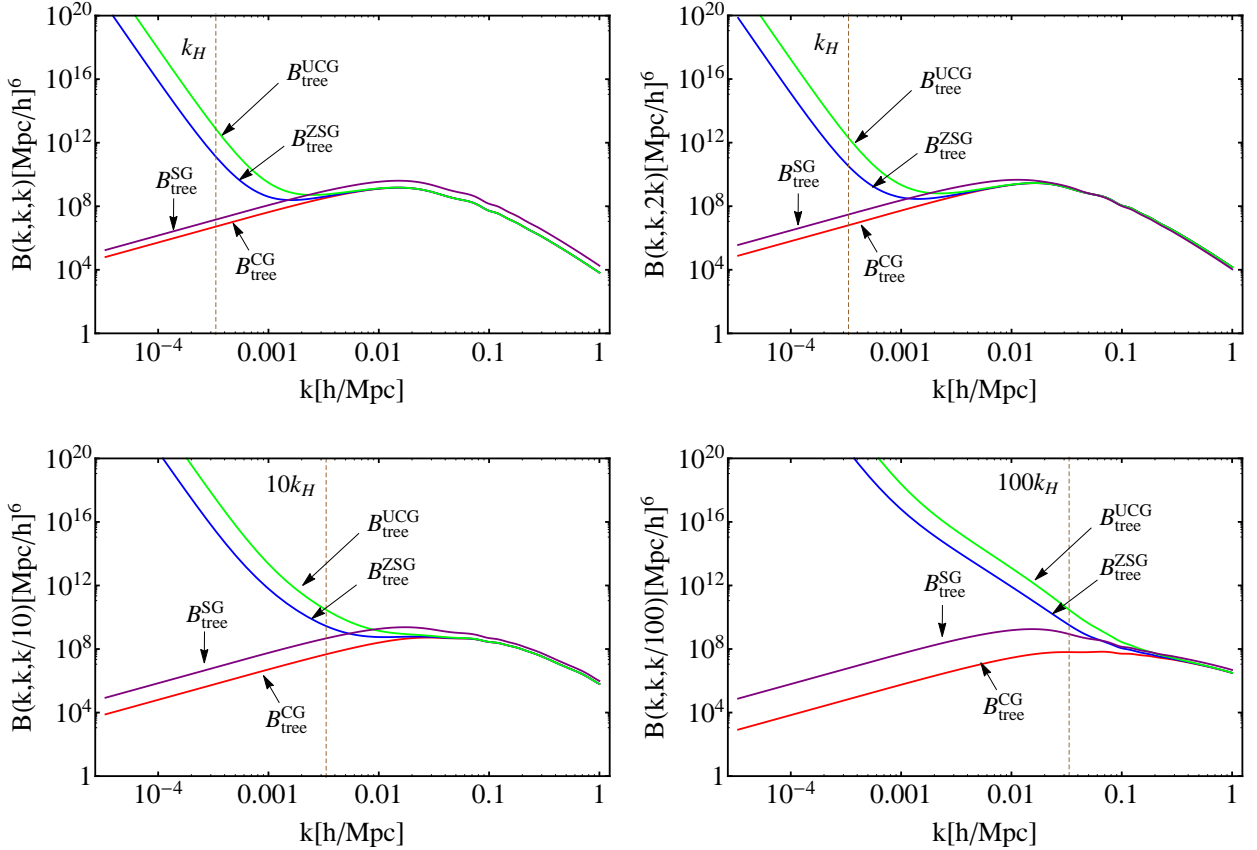}
  \caption{
 We show the tree bispectrum in the comoving (CG), synchronous (SG), zero shear (ZSG) and uniform curvature (UCG) gauges. From top left clockwise, the bispectra are projected onto the equilateral, folded, tightly ($\alpha=100$) and slightly ($\alpha=10$) squeezed configurations at $z=0$.}
  \label{fig:gauges}
\end{figure}

We compare the tree level bispectra in all the aforementioned gauges
in Figure~\ref{fig:gauges}.
On small scales, the tree-level bispectra from all gauges except for
the synchronous gauge converge to the Newtonian tree level bispectrum:
the bispectrum in the synchronous gauge does not converge to the Newtonian
(Eulerian) bispectrum, because the coordinate system is the similar
to the Lagrangian fluid view.
On larger scales ($k_3 \lesssim k_H$), we start to see the
gauge dependence of the matter density fields and the tree level
bispectra from all four gauges are different from each other.

\section{Conclusions}
\label{sec:conclusion}

We have studied how the non-linearities in general relativistic
affects the non-linear density and velocity bispectra.
Using the full general relativistic formalism, we calculate one-loop
bispectra of density and velocity fields
in a flat, matter
dominated universe. We have assumed that the initial density
perturbation is perfectly Gaussian, so that
the matter bispectrum comes solely from the non-linear dynamics.
As we work in the comoving gauge, where the relativistic
density and velocity fields coincide with those in the Newtonian theory, the
pure relativistic corrections appear from third order. We have computed
the non-linear bispectrum in the equilateral, folded and squeezed triangular
configurations and have shown that the relativistic effects are appreciable
only on the scale larger than the horizon.
On small scales, the Newtonian one-loop corrections dominate the relativistic
ones and even the tree contributions at $k\gtrsim0.2h$Mpc${}^{-1}$,
indicating the non-linear evolution of the bispectrum due to gravitational
instability.
The general relativistic corrections appear to be dominant over the Newtonian
ones when the longest wavemode is near comoving horizon scale.
That is the reason why we see the domination of the relativistic corrections
in the squeezed configurations on very large scales.

We then have demonstrated the gauge dependence of the matter bispectrum by
explicitly computing the tree-level matter bispectrum in
four different gauges: comoving, synchronous, zero shear and uniform curvature gauges.
Except for the synchronous gauge, whose time coordinate is comoving with the
observer and thus the meaning of the density contrast differs from the other
gauges even on the small scales, the matter bispectrum computed in the other
gauges are the same on small scales.
On large scales, the gauge dependence begin to show up more prominently and
the matter bisepctra from all four gauges diverge from each other.
This result is, again, consistent with the one-loop result in the comoving
gauge that the general relativistic effects are only important near the
horizon scales.

The gauge dependence that we see near the horizon scale is the outcome of
that the matter bispectrum itself is not a direct observable.
When considering the observable quantities such as the bispectrum of
weak-lensing shear and convergence field, or the galaxy bispectrum including
all the relevant effects such as galaxy bias, light reflection, etc, the gauge
dependence must disappear. In the case of the galaxy power spectrum,
\cite{gaugedep} have shown that the observed galaxy power spectrum written
in terms of the observed coordinate system is indeed gauge independent.
Likewise, we surmise that calculating the bispectrum of observed
quantities should resolve this gauge dependent ambiguities.
Such a calculation requires extending the previous work done for the galaxy
power spectrum to second order.

\subsection*{Acknowledgements}

SGB appreciates Jai-chan Hwang for suggesting helpful comments and supporting this research.
We thank the Topical Research Program ``Theories and practices in large scale structure formation'', supported by the Asia Pacific Center for Theoretical Physics, while this work was under progress.
JG acknowledges the Max-Planck-Gesellschaft, the Korea Ministry of Education, Science and Technology, Gyeongsangbuk-Do and Pohang City for
the support of the Independent Junior Research Group at the Asia Pacific Center for Theoretical Physics.
JG is also supported by a Starting Grant through the Basic Science Research Program of the National Research Foundation of Korea (2013R1A1A1006701).
DJ is supported by DoE SC-0008108 and NASA NNX12AE86G, and acknowledges support from the John Templeton Foundation.

\appendix

\section{Fourth order solutions}
\label{app:sol}

Substituting (\ref{nldelta}) and (\ref{nltheta}) into (\ref{continuity}) and (\ref{euler}), in Fourier space we obtain the differential equations for the fourth order kernels $F_4$ and $G_4$ as
\begin{align}
& \frac{1}{H}\frac{dF_4}{dt} + 4F_4 - G_4
\nonumber\\
=&  \frac{{\bm k}\cdot{\bm q}_{234}}{q_{234}^2}G_3({\bm q}_2,{\bm q}_3,{\bm q}_4) + \frac{{\bm k}\cdot{\bm q}_{34}}{q_{34}^2}F_2({\bm q}_1,{\bm q}_2)G_2({\bm q}_3,{\bm q}_4) + \frac{{\bm k}\cdot{\bm q}_1}{q_1^2}F_3({\bm q}_2,{\bm q}_3,{\bm q}_4)
\nonumber\\
& - (aH)^2 \bigg\{ \frac{5}{2} \bigg[ \frac{{\bm q}_{12}\cdot{\bm q}_{34}}{q_{12}^2} \bigg( -\frac{2}{q_2^2} + \frac{{\bm q}_1\cdot{\bm q}_2}{q_1^2q_2^2} - \frac{3}{2}\frac{{\bm q}_{12}\cdot{\bm q}_1}{q_{12}^2q_1^2} - \frac{3}{2}\frac{{\bm q}_{12}\cdot{\bm q}_2}{q_{12}^2}\frac{{\bm q}_1\cdot{\bm q}_2}{q_1^2q_2^2} \bigg) + 2\frac{{\bm q}_1\cdot{\bm q}_{34}}{q_1^2q_2^2} \bigg] F_2({\bm q}_3,{\bm q}_4)
\nonumber\\
& \hspace{0.6cm} + \bigg[ \frac{{\bm q}_{123}\cdot{\bm q}_4}{q_{123}^2} \bigg( -\frac{2}{q_{23}^2} + \frac{{\bm q}_1\cdot{\bm q}_{23}}{q_1^2q_{23}^2} - \frac{3}{2}\frac{{\bm q}_{123}\cdot{\bm q}_1}{q_{123}^2q_1^2} - \frac{3}{2}\frac{{\bm q}_{123}\cdot{\bm q}_{23}}{q_{123}^2}\frac{{\bm q}_1\cdot{\bm q}_{23}}{q_1^2q_{23}^2} \bigg) + 2\frac{{\bm q}_1\cdot{\bm q}_4}{q_1^2q_{23}^2} \bigg] \bigg[ \frac{3}{2}F_2({\bm q}_2,{\bm q}_3) + G_2({\bm q}_2,{\bm q}_3) \bigg]
\nonumber\\
& \hspace{0.6cm} + \frac{5}{2} \bigg[ \frac{{\bm q}_{123}\cdot{\bm q}_4}{q_{123}^2} \bigg( -\frac{2}{q_1^2} + \frac{{\bm q}_1\cdot{\bm q}_{23}}{q_1^2q_{23}^2} - \frac{3}{2}\frac{{\bm q}_{123}\cdot{\bm q}_{23}}{q_{123}^2q_{23}^2} - \frac{3}{2}\frac{{\bm q}_{123}\cdot{\bm q}_1}{q_{123}^2}\frac{{\bm q}_1\cdot{\bm q}_{23}}{q_1^2q_{23}^2} \bigg) + 2\frac{{\bm q}_{23}\cdot{\bm q}_4}{q_1^2q_{23}^2} \bigg] G_2({\bm q}_2,{\bm q}_3) \bigg\}
\nonumber\\
& + \frac{{\bm q}_1\cdot{\bm q}_{234}}{q_{234}^2} \bigg\{ \frac{25}{4}(aH)^4 \bigg( \frac{2}{q_4^2} - \frac{{\bm q}_{23}\cdot{\bm q}_4}{q_{23}^2q_4^2} + \frac{3}{2}\frac{{\bm q}_{234}\cdot{\bm q}_{23}}{q_{234}^2q_{23}^2} + \frac{3}{2}\frac{{\bm q}_{234}\cdot{\bm q}_4}{q_{234}^2}\frac{{\bm q}_{23}\cdot{\bm q}_4}{q_{23}^2q_4^2} \bigg)
& \hspace{3.3cm} \times
\bigg( \frac{2}{q_2^2} - \frac{{\bm q}_2\cdot{\bm q}_3}{q_2^2q_3^2} + \frac{3}{2}\frac{{\bm q}_{23}\cdot{\bm q}_2}{q_{23}^2q_2^2} + \frac{3}{2}\frac{{\bm q}_{23}\cdot{\bm q}_3}{q_{23}^2}\frac{{\bm q}_2\cdot{\bm q}_3}{q_2^2q_3^2} \bigg)
\nonumber\\
& \hspace{0.6cm} + \frac{(aH)^2}{2q_{34}^2} \bigg( -2 + \frac{{\bm q}_2\cdot{\bm q}_{34}}{q_2^2} - \frac{3}{2}\frac{{\bm q}_{234}\cdot{\bm q}_2}{q_{234}^2}\frac{q_{34}^2}{q_2^2} - \frac{3}{2}\frac{{\bm q}_{234}\cdot{\bm q}_{34}}{q_{234}^2}\frac{{\bm q}_2\cdot{\bm q}_{34}}{q_{34}^2} \bigg)
\nonumber\\
& \hspace{1.8cm} \times \bigg\{ \frac{1}{2} \bigg[ 1 + \frac{{\bm q}_3\cdot{\bm q}_4}{q_3^2} \bigg( 1 - \frac{{\bm q}_{34}\cdot{\bm q}_4}{q_4^2} \bigg) \bigg] - \frac{25}{4} \frac{(aH)^2}{q_3^2q_4^2} \left( 3{\bm q}_3\cdot{\bm q}_4 + 8q_4^2 \right) \bigg\}
\nonumber\\
& \hspace{0.6cm} + \frac{25}{4}\frac{(aH)^4}{q_3^2q_4^2} \bigg[ -4 + 4\frac{{\bm q}_2\cdot{\bm q}_3}{q_2^2} - \frac{3}{2} \bigg( \frac{{\bm q}_{234}\cdot{\bm q}_2}{q_{234}^2}\frac{{\bm q}_3\cdot{\bm q}_4}{q_2^2} + 3\frac{{\bm q}_{234}\cdot{\bm q}_4}{q_{234}^2}\frac{{\bm q}_2\cdot{\bm q}_3}{q_2^2}
+ 4\frac{{\bm q}_{234}\cdot{\bm q}_3}{q_{234}^2}\frac{{\bm q}_2\cdot{\bm q}_3}{q_2^2} + 4\frac{{\bm q}_{234}\cdot{\bm q}_2}{q_{234}^2}\frac{q_3^2}{q_2^2} \bigg) \bigg] \bigg\}
\nonumber\\
& + 2 \bigg\{ \frac{{\bm q}_1\cdot{\bm q}_2}{2q_2^2q_{34}^2}
\bigg( \frac{(aH)^2}{2} \bigg[ 1 + \frac{{\bm q}_3\cdot{\bm q}_4}{q_3^2} \bigg( 1 - \frac{{\bm q}_{34}\cdot{\bm q}_4}{q_4^2} \bigg) \bigg] - \frac{25}{4}\frac{(aH)^4}{q_3^2q_4^2} \left( 3{\bm q}_3\cdot{\bm q}_4 + 8q_4^2 \right) \bigg)
\nonumber\\
& \hspace{0.6cm} + \frac{25}{4}\frac{(aH)^4}{q_2^2}\frac{{\bm q}_1\cdot{\bm q}_{34}}{q_{34}^2} \bigg[ -\frac{2}{q_4^2} + \frac{{\bm q}_3\cdot{\bm q}_4}{q_3^2q_4^2} - \frac{3}{2}\frac{{\bm q}_{34}\cdot{\bm q}_3}{q_{34}^2q_3^2} - \frac{3}{2}\frac{{\bm q}_{34}\cdot{\bm q}_4}{q_{34}^2}\frac{{\bm q}_3\cdot{\bm q}_4}{q_3^2q_4^2} \bigg] \bigg\} - 25\frac{{\bm q}_1\cdot{\bm q}_2}{q_2^2}\frac{(aH)^4}{q_3^2q_4^2} \, ,
\end{align}
and
\begin{align}
& \frac{1}{H}\frac{dG_4}{dt} + \frac{3}{2} \left( 3G_4 - F_4 \right)
\nonumber\\
= & \frac{k^2{\bm q}_1\cdot{\bm q}_{234}}{q_1^2q_{234}^2}G_3({\bm q}_2,{\bm q}_3,{\bm q}_4) + \frac{k^2{\bm q}_{12}\cdot{\bm q}_{34}}{2q_{12}^2q_{34}^2}G_2({\bm q}_1,{\bm q}_2)G_2({\bm q}_3,{\bm q}_4)
\nonumber\\
& + (aH)^2 \bigg( \bigg\{ \bigg[ \frac{2}{3} + \frac{{\bm q}_1\cdot{\bm q}_{234}}{q_1^2} \bigg( 1 - \frac{k^2}{q_{234}^2} \bigg) \bigg] \bigg( -\frac{2}{q_2^2} + \frac{{\bm q}_2\cdot{\bm q}_{34}}{q_2^2q_{34}^2} - \frac{3}{2}\frac{{\bm q}_{234}\cdot{\bm q}_{34}}{q_{234}^2q_{34}^2} - \frac{3}{2}\frac{{\bm q}_{234}\cdot{\bm q}_2}{q_{234}^2}\frac{{\bm q}_2\cdot{\bm q}_{34}}{q_2^2q_{34}^2} \bigg)
\nonumber\\
& \hspace{0.6cm} + \frac{1}{q_2^2} \bigg[ \frac{2}{3}\frac{{\bm q}_1\cdot{\bm q}_{34}}{q_{34}^2} - 4 \bigg( \frac{{\bm q}_1\cdot{\bm q}_{34}}{q_{34}^2} - \frac{1}{3} \bigg) \frac{{\bm k}\cdot{\bm q}_1}{q_1^2} \bigg] \bigg\} \frac{5}{2}G_2({\bm q}_3,{\bm q}_4)
\nonumber\\
& \hspace{1.5cm} + \bigg\{ \bigg[ \frac{2}{3} + \frac{{\bm q}_{12}\cdot{\bm q}_{34}}{q_{34}^2} \bigg( 1 - \frac{k^2}{q_{12}^2} \bigg) \bigg] \bigg( -\frac{2}{q_2^2} + \frac{{\bm q}_1\cdot{\bm q}_2}{q_1^2q_2^2} - \frac{3}{2}\frac{{\bm q}_{12}\cdot{\bm q}_1}{q_{12}^2q_1^2} - \frac{3}{2}\frac{{\bm q}_{12}\cdot{\bm q}_2}{q_{12}^2}\frac{{\bm q}_1\cdot{\bm q}_2}{q_1^2q_2^2} \bigg)
\nonumber\\
& \hspace{2cm} + \frac{1}{q_2^2} \bigg[ \frac{2}{3}\frac{{\bm q}_1\cdot{\bm q}_{34}}{q_1^2} - 4 \bigg( \frac{{\bm q}_1\cdot{\bm q}_{34}}{q_1^2} - \frac{1}{3} \bigg) \frac{{\bm k}\cdot{\bm q}_{34}}{q_{34}^2} \bigg] \bigg\} \frac{5}{2}G_2({\bm q}_3,{\bm q}_4)
\nonumber\\
& \hspace{1.5cm} + \bigg\{ \bigg[ \frac{2}{3} + \frac{{\bm q}_1\cdot{\bm q}_{234}}{q_1^2} \bigg( 1 - \frac{k^2}{q_{234}^2} \bigg) \bigg] \bigg( -\frac{2}{q_{34}^2} + \frac{{\bm q}_2\cdot{\bm q}_{34}}{q_2^2q_{34}^2} - \frac{3}{2}\frac{{\bm q}_2\cdot{\bm q}_{234}}{q_2^2q_{234}^2} - \frac{3}{2}\frac{{\bm q}_{234}\cdot{\bm q}_{34}}{q_{234}^2}\frac{{\bm q}_2\cdot{\bm q}_{34}}{q_2^2q_{34}^2} \bigg)
\nonumber\\
& \hspace{2cm} + \frac{1}{q_{34}^2} \bigg[ \frac{2}{3}\frac{{\bm q}_1\cdot{\bm q}_2}{q_2^2} - 4 \bigg( \frac{{\bm q}_1\cdot{\bm q}_2}{q_2^2} - \frac{1}{3} \bigg) \frac{{\bm k}\cdot{\bm q}_1}{q_1^2} \bigg] \bigg\} \bigg[ \frac{3}{2}F_2({\bm q}_3,{\bm q}_4) + G_2({\bm q}_3,{\bm q}_4) \bigg] \bigg)
\nonumber\\
& + \bigg( -\frac{2}{3} - \frac{{\bm q}_1\cdot{\bm q}_{234}}{q_1^2} + \frac{k^2}{q_{234}^2}\frac{{\bm q}_1\cdot{\bm q}_{234}}{q_1^2} \bigg)
\nonumber\\
& \hspace{0.3cm} \times \bigg\{ \frac{25}{4} (aH)^4 \bigg( \frac{2}{q_4^2} - \frac{{\bm q}_{23}\cdot{\bm q}_4}{q_{23}^2q_4^2} + \frac{3}{2}\frac{{\bm q}_{234}\cdot{\bm q}_{23}}{q_{234}^2q_{23}^2} + \frac{3}{2}\frac{{\bm q}_{234}\cdot{\bm q}_4}{q_{234}^2}\frac{{\bm q}_{23}\cdot{\bm q}_4}{q_{23}^2q_4^2} \bigg)\nonumber\\
& \hspace{2.2cm}\times\bigg( \frac{2}{q_3^2} - \frac{{\bm q}_2\cdot{\bm q}_3}{q_2^2q_3^2} + \frac{3}{2}\frac{{\bm q}_{23}\cdot{\bm q}_2}{q_{23}^2q_2^2} + \frac{3}{2}\frac{{\bm q}_{23}\cdot{\bm q}_3}{q_{23}^2}\frac{{\bm q}_2\cdot{\bm q}_3}{q_2^2q_3^2} \bigg)
\nonumber\\
& \hspace{0.8cm} + \frac{(aH)^2}{2q_{34}^2} \bigg( -2 + \frac{{\bm q}_2\cdot{\bm q}_{34}}{q_2^2} - \frac{3}{2}\frac{{\bm q}_{234}\cdot{\bm q}_2}{q_{234}^2}\frac{q_{34}^2}{q_2^2} - \frac{3}{2}\frac{{\bm q}_{234}\cdot{\bm q}_{34}}{q_{234}^2}\frac{{\bm q}_2\cdot{\bm q}_{34}}{q_{34}^2} \bigg)
\nonumber\\
& \hspace{2.3cm} \times \bigg\{ \frac{1}{2} \bigg[ 1 + \frac{{\bm q}_3\cdot{\bm q}_4}{q_3^2} \bigg( 1 - \frac{{\bm q}_{34}\cdot{\bm q}_4}{q_4^2} \bigg) \bigg] - \frac{25}{4}\frac{(aH)^2}{q_3^2q_4^2} \left( 3{\bm q}_3\cdot{\bm q}_4 + 8q_4^2 \right) \bigg\}
\nonumber\\
& \hspace{0.8cm} + \frac{25}{4}\frac{(aH)^4}{q_3^2q_4^2} \bigg[ -4 + 4\frac{{\bm q}_2\cdot{\bm q}_3}{q_2^2} - \frac{3}{2} \bigg( \frac{{\bm q}_{234}\cdot{\bm q}_2}{q_{234}^2}\frac{{\bm q}_3\cdot{\bm q}_4}{q_2^2} + 3\frac{{\bm q}_{234}\cdot{\bm q}_4}{q_{234}^2}\frac{{\bm q}_2\cdot{\bm q}_3}{q_2^2} + 4\frac{{\bm q}_{234}\cdot{\bm q}_3}{q_{234}^2}\frac{{\bm q}_2\cdot{\bm q}_3}{q_2^2} + 4\frac{{\bm q}_{234}\cdot{\bm q}_2}{q_{234}^2}\frac{q_3^2}{q_2^2} \bigg) \bigg] \bigg\}
\nonumber\\
&+  \frac{(aH)^2}{2q_{34}^2} \bigg( -\frac{2}{3}\frac{{\bm q}_1\cdot{\bm q}_2}{q_1^2} + 4\frac{{\bm q}_1\cdot{\bm q}_2}{q_1^2}\frac{{\bm k}\cdot{\bm q}_2}{q_2^2} - \frac{4}{3}\frac{{\bm k}\cdot{\bm q}_2}{q_2^2} \bigg)\nonumber\\
&\hspace{2cm}\times\bigg\{ \frac{1}{2} \bigg[ 1 + \frac{{\bm q}_3\cdot{\bm q}_4}{q_3^2} \bigg( 1 - \frac{{\bm q}_{34}\cdot{\bm q}_4}{q_4^2} \bigg) \bigg] - \frac{25}{4}\frac{(aH)^2}{q_3^2q_4^2} \left( 3{\bm q}_3\cdot{\bm q}_4 + 8q_4^2 \right) \bigg\}
\nonumber\\
& + \frac{25}{4}\frac{(aH)^4}{q_1^2} \bigg( -\frac{2}{q_4^2} + \frac{{\bm q}_3\cdot{\bm q}_4}{q_3^2q_4^2} - \frac{3}{2}\frac{{\bm q}_{34}\cdot{\bm q}_3}{q_{34}^2q_3^2} - \frac{3}{2}\frac{{\bm q}_{34}\cdot{\bm q}_4}{q_{34}^2}\frac{{\bm q}_3\cdot{\bm q}_4}{q_3^2q_4^2} \bigg)
\nonumber\\
& \hspace{2cm} \times \bigg( 4\frac{{\bm k}\cdot{\bm q}_2}{q_2^2}\frac{{\bm q}_2\cdot{\bm q}_{34}}{q_{34}^2} - \frac{4}{3}\frac{{\bm k}\cdot{\bm q}_2}{q_2^2} + 4\frac{{\bm k}\cdot{\bm q}_{34}}{q_{34}^2}\frac{{\bm q}_2\cdot{\bm q}_{34}}{q_2^2} - \frac{4}{3}\frac{{\bm k}\cdot{\bm q}_{34}}{q_{34}^2} - \frac{2}{3}\frac{{\bm q}_2\cdot{\bm q}_{34}}{q_{34}^2} - \frac{8}{3}\frac{{\bm q}_2\cdot{\bm q}_{34}}{{q_2}^2} \bigg)
\nonumber\\
& + \frac{25}{4}\frac{(aH)^4}{q_1^2q_2^2} \bigg( 12\frac{{\bm q}_3\cdot{\bm q}_4}{q_3^2}\frac{{\bm k}\cdot{\bm q}_4}{q_4^2} - 4\frac{{\bm k}\cdot{\bm q}_4}{q_4^2} - 4\frac{{\bm q}_3\cdot{\bm q}_4}{q_3^2} + 2\frac{{\bm q}_1\cdot{\bm q}_2}{q_3^2}\frac{{\bm q}_3\cdot{\bm q}_4}{q_4^2} + \frac{2}{3}\frac{{\bm q}_1\cdot{\bm q}_3}{q_3^2}\frac{{\bm q}_2\cdot{\bm q}_4}{q_4^2} \bigg)
\nonumber\\
& + \frac{25}{4}(aH)^4 \bigg( \frac{2}{3} - \frac{{\bm k}\cdot{\bm q}_{34}}{q_{34}^2} + \frac{{\bm k}\cdot{\bm q}_{34}}{q_{34}^2}\frac{{\bm q}_{12}\cdot{\bm q}_{34}}{q_{12}^2} \bigg)
\nonumber\\
& \hspace{2cm} \times \bigg( \frac{2}{q_2^2} - \frac{{\bm q}_1\cdot{\bm q}_2}{q_1^2q_2^2} + \frac{3}{2}\frac{{\bm q}_{12}\cdot{\bm q}_1}{q_{12}^2q_1^2} + \frac{3}{2}\frac{{\bm q}_{12}\cdot{\bm q}_2}{q_{12}^2}\frac{{\bm q}_1\cdot{\bm q}_2}{q_1^2q_2^2} \bigg) \bigg( \frac{2}{q_4^2} - \frac{{\bm q}_3\cdot{\bm q}_4}{q_3^2q_4^2} + \frac{3}{2}\frac{{\bm q}_{34}\cdot{\bm q}_3}{q_{34}^2q_3^2} + \frac{3}{2}\frac{{\bm q}_{34}\cdot{\bm q}_4}{q_{34}^2}\frac{{\bm q}_3\cdot{\bm q}_4}{q_3^2q_4^2} \bigg) \, .
\end{align}

We can obtain readily the solutions of these equations. We can divide the kernels into the Newtonian and general relativistic parts. The Newtonian kernels are time independent, and can be calculated algebraically in terms of lower order kernels as
\begin{align}
F_4^\text{(N)} = & \frac{1}{33} \bigg[ \frac{9{\bm k}\cdot{\bm q}_1}{q_1^2}F_3^\text{(N)}({\bm q}_2,{\bm q}_3,{\bm q}_4) + \frac{9{\bm k}\cdot{\bm q}_{234}}{q_{234}^2}G_3^\text{(N)}({\bm q}_2,{\bm q}_3,{\bm q}_4) + \frac{9{\bm k}\cdot{\bm q}_{34}}{q_{34}^2}F_2({\bm q}_1,{\bm q}_2)G_2({\bm q}_3,{\bm q}_4)
\nonumber\\
& \hspace{0.5cm} + \frac{2k^2{\bm q}_1\cdot{\bm q}_{234}}{q_1^2q_{234}^2}G_3^\text{(N)}({\bm q}_2,{\bm q}_3,{\bm q}_4) + \frac{2k^2{\bm q}_{12}\cdot{\bm q}_{34}}{2q_{12}^2q_{34}^2}G_2({\bm q}_1,{\bm q}_2)G_2({\bm q}_3,{\bm q}_4) \bigg] \, ,
\\
G_4^\text{(N)} = & \frac{1}{33} \bigg[ \frac{3{\bm k}\cdot{\bm q}_1}{q_1^2}F_3^\text{(N)}({\bm q}_2,{\bm q}_3,{\bm q}_4) + \frac{3{\bm k}\cdot{\bm q}_{234}}{q_{234}^2}G_3^\text{(N)}({\bm q}_2,{\bm q}_3,{\bm q}_4) + \frac{3{\bm k}\cdot{\bm q}_{34}}{q_{34}^2}F_2({\bm q}_1,{\bm q}_2)G_2({\bm q}_3,{\bm q}_4)
\nonumber\\
& \hspace{0.5cm} + \frac{8k^2{\bm q}_1\cdot{\bm q}_{234}}{q_1^2q_{234}^2}G_3^\text{(N)}({\bm q}_2,{\bm q}_3,{\bm q}_4) + \frac{8k^2{\bm q}_{12}\cdot{\bm q}_{34}}{2q_{12}^2q_{34}^2}G_2({\bm q}_1,{\bm q}_2)G_2({\bm q}_3,{\bm q}_4) \bigg] \, .
\end{align}
Meanwhile, the general relativistic kernels are time dependent. Introducing the horizon scale wavenumber $k_H \equiv aH$, we can find
\begin{align}
F_4^\text{(GR)} & = \frac{7A_1+2B_1}{18} k_H^2 + \frac{5A_2+2B_2}{7}k_H^4  \, ,
\\
G_4^\text{(GR)} & = \frac{A_1+2B_1}{6}k_H^2 + \frac{3A_2+4B_2}{7}k_H^4  \, ,
\end{align}
where
\begin{align}
& A_1 = \frac{{\bm k}\cdot{\bm q}_1}{q_1^2}k_H^{-2}F_3^\text{(GR)}({\bm q}_2,{\bm q}_3,{\bm q}_4) + \frac{{\bm k}\cdot{\bm q}_{234}}{q_{234}^2}k_H^{-2}G_3^\text{(GR)}({\bm q}_2,{\bm q}_3,{\bm q}_4)
\nonumber\\
& - \frac{5}{2} \bigg[ \frac{{\bm q}_{12}\cdot{\bm q}_{34}}{q_{12}^2} \bigg( -\frac{2}{q_2^2} + \frac{{\bm q}_1\cdot{\bm q}_2}{q_1^2q_2^2} - \frac{3}{2}\frac{{\bm q}_{12}\cdot{\bm q}_1}{q_{12}^2q_1^2} - \frac{3}{2}\frac{{\bm q}_{12}\cdot{\bm q}_2}{q_{12}^2}\frac{{\bm q}_1\cdot{\bm q}_2}{q_1^2q_2^2} \bigg) + 2\frac{{\bm q}_1\cdot{\bm q}_{34}}{q_1^2q_2^2} \bigg] F_2({\bm q}_3,{\bm q}_4)
\nonumber\\
& - \bigg[ \frac{{\bm q}_{123}\cdot{\bm q}_4}{q_{123}^2} \bigg( -\frac{2}{q_{23}^2} + \frac{{\bm q}_1\cdot{\bm q}_{23}}{q_1^2q_{23}^2} - \frac{3}{2}\frac{{\bm q}_{123}\cdot{\bm q}_1}{q_{123}^2q_1^2} - \frac{3}{2}\frac{{\bm q}_{123}\cdot{\bm q}_{23}}{q_{123}^2}\frac{{\bm q}_1\cdot{\bm q}_{23}}{q_1^2q_{23}^2} \bigg) + 2\frac{{\bm q}_1\cdot{\bm q}_4}{q_1^2q_{23}^2} \bigg] \bigg[ \frac{3}{2}F_2({\bm q}_2,{\bm q}_3) + G_2({\bm q}_2,{\bm q}_3) \bigg]
\nonumber\\
& - \frac{5}{2} \bigg[ \frac{{\bm q}_{123}\cdot{\bm q}_4}{q_{123}^2} \bigg( -\frac{2}{q_1^2} + \frac{{\bm q}_1\cdot{\bm q}_{23}}{q_1^2q_{23}^2} - \frac{3}{2}\frac{{\bm q}_{123}\cdot{\bm q}_{23}}{q_{123}^2q_{23}^2} - \frac{3}{2}\frac{{\bm q}_{123}\cdot{\bm q}_1}{q_{123}^2}\frac{{\bm q}_1\cdot{\bm q}_{23}}{q_1^2q_{23}^2} \bigg) + 2\frac{{\bm q}_{23}\cdot{\bm q}_4}{q_1^2q_{23}^2} \bigg] G_2({\bm q}_2,{\bm q}_3)
\nonumber\\
& + \frac{{\bm q}_1\cdot{\bm q}_{234}}{q_{234}^2} \frac{1}{4q_{34}^2} \bigg( -2 + \frac{{\bm q}_2\cdot{\bm q}_{34}}{q_2^2} - \frac{3}{2}\frac{{\bm q}_{234}\cdot{\bm q}_2}{q_{234}^2}\frac{q_{34}^2}{q_2^2} - \frac{3}{2}\frac{{\bm q}_{234}\cdot{\bm q}_{34}}{q_{234}^2}\frac{{\bm q}_2\cdot{\bm q}_{34}}{q_{34}^2} \bigg) \bigg[ 1 + \frac{{\bm q}_3\cdot{\bm q}_4}{q_3^2} \bigg( 1 - \frac{{\bm q}_{34}\cdot{\bm q}_4}{q_4^2} \bigg) \bigg]
\nonumber\\
& + \frac{{\bm q}_1\cdot{\bm q}_2}{2q_2^2q_{34}^2} \bigg[ 1 + \frac{{\bm q}_3\cdot{\bm q}_4}{q_3^2} \bigg( 1 - \frac{{\bm q}_{34}\cdot{\bm q}_4}{q_4^2} \bigg) \bigg] \, ,
\\
& B_1 = \frac{k^2{\bm q}_1\cdot{\bm q}_{234}}{q_1^2q_{234}^2}k_H^{-2}G_3^\text{(GR)}({\bm q}_2,{\bm q}_3,{\bm q}_4)
\nonumber\\
& + \frac{5}{2}\bigg\{ \bigg[ \frac{2}{3} + \frac{{\bm q}_1\cdot{\bm q}_{234}}{q_1^2} \bigg( 1 - \frac{k^2}{q_{234}^2} \bigg) \bigg] \bigg( -\frac{2}{q_2^2} + \frac{{\bm q}_2\cdot{\bm q}_{34}}{q_2^2q_{34}^2} - \frac{3}{2}\frac{{\bm q}_{234}\cdot{\bm q}_{34}}{q_{234}^2q_{34}^2} - \frac{3}{2}\frac{{\bm q}_{234}\cdot{\bm q}_2}{q_{234}^2}\frac{{\bm q}_2\cdot{\bm q}_{34}}{q_2^2q_{34}^2} \bigg)
\nonumber\\
& \hspace{0.8cm} + \frac{1}{q_2^2} \bigg[ \frac{2}{3}\frac{{\bm q}_1\cdot{\bm q}_{34}}{q_{34}^2} - 4 \bigg( \frac{{\bm q}_1\cdot{\bm q}_{34}}{q_{34}^2} - \frac{1}{3} \bigg) \frac{{\bm k}\cdot{\bm q}_1}{q_1^2} \bigg] \bigg\} G_2({\bm q}_3,{\bm q}_4)
\nonumber\\
& + \frac{5}{2} \bigg\{ \bigg[ \frac{2}{3} + \frac{{\bm q}_{12}\cdot{\bm q}_{34}}{q_{34}^2} \bigg( 1 - \frac{k^2}{q_{12}^2} \bigg) \bigg] \bigg( -\frac{2}{q_2^2} + \frac{{\bm q}_1\cdot{\bm q}_2}{q_1^2q_2^2} - \frac{3}{2}\frac{{\bm q}_{12}\cdot{\bm q}_1}{q_{12}^2q_1^2} - \frac{3}{2}\frac{{\bm q}_{12}\cdot{\bm q}_2}{q_{12}^2}\frac{{\bm q}_1\cdot{\bm q}_2}{q_1^2q_2^2} \bigg)
\nonumber\\
& \hspace{0.8cm} + \frac{1}{q_2^2} \bigg[ \frac{2}{3}\frac{{\bm q}_1\cdot{\bm q}_{34}}{q_1^2} - 4 \bigg( \frac{{\bm q}_1\cdot{\bm q}_{34}}{q_1^2} - \frac{1}{3} \bigg) \frac{{\bm k}\cdot{\bm q}_{34}}{q_{34}^2} \bigg] \bigg\} G_2({\bm q}_3,{\bm q}_4)
\nonumber\\
& + \bigg\{ \bigg[ \frac{2}{3} + \frac{{\bm q}_1\cdot{\bm q}_{234}}{q_1^2} \bigg( 1 - \frac{k^2}{q_{234}^2} \bigg) \bigg] \bigg( -\frac{2}{q_{34}^2} + \frac{{\bm q}_2\cdot{\bm q}_{34}}{q_2^2q_{34}^2} - \frac{3}{2}\frac{{\bm q}_2\cdot{\bm q}_{234}}{q_2^2q_{234}^2} - \frac{3}{2}\frac{{\bm q}_{234}\cdot{\bm q}_{34}}{q_{234}^2}\frac{{\bm q}_2\cdot{\bm q}_{34}}{q_2^2q_{34}^2} \bigg)
\nonumber\\
& \hspace{0.8cm} + \frac{1}{q_{34}^2} \bigg[ \frac{2}{3}\frac{{\bm q}_1\cdot{\bm q}_2}{q_2^2} - 4 \bigg( \frac{{\bm q}_1\cdot{\bm q}_2}{q_2^2} - \frac{1}{3} \bigg) \frac{{\bm k}\cdot{\bm q}_1}{q_1^2} \bigg] \bigg\} \bigg[ \frac{3}{2}F_2({\bm q}_3,{\bm q}_4) + G_2({\bm q}_3,{\bm q}_4) \bigg]
\nonumber\\
& + \bigg( -\frac{2}{3} - \frac{{\bm q}_1\cdot{\bm q}_{234}}{q_1^2} + \frac{k^2}{q_{234}^2}\frac{{\bm q}_1\cdot{\bm q}_{234}}{q_1^2} \bigg)
\nonumber\\
& \hspace{0.5cm} \times \frac{1}{4q_{34}^2} \bigg( -2 + \frac{{\bm q}_2\cdot{\bm q}_{34}}{q_2^2} - \frac{3}{2}\frac{{\bm q}_{234}\cdot{\bm q}_2}{q_{234}^2}\frac{q_{34}^2}{q_2^2} - \frac{3}{2}\frac{{\bm q}_{234}\cdot{\bm q}_{34}}{q_{234}^2}\frac{{\bm q}_2\cdot{\bm q}_{34}}{q_{34}^2} \bigg) \bigg[ 1 + \frac{{\bm q}_3\cdot{\bm q}_4}{q_3^2} \bigg( 1 - \frac{{\bm q}_{34}\cdot{\bm q}_4}{q_4^2} \bigg) \bigg]
\nonumber\\
& + \frac{1}{4q_{34}^2} \bigg( -\frac{2}{3}\frac{{\bm q}_1\cdot{\bm q}_2}{q_1^2} + 4\frac{{\bm q}_1\cdot{\bm q}_2}{q_1^2}\frac{{\bm k}\cdot{\bm q}_2}{q_2^2} - \frac{4}{3}\frac{{\bm k}\cdot{\bm q}_2}{q_2^2} \bigg) \bigg[ 1 + \frac{{\bm q}_3\cdot{\bm q}_4}{q_3^2} \bigg( 1 - \frac{{\bm q}_{34}\cdot{\bm q}_4}{q_4^2} \bigg) \bigg] \, ,
\end{align}
and
\begin{align}
& A_2 = \frac{25}{4} \frac{{\bm q}_1\cdot{\bm q}_{234}}{q_{234}^2} \bigg( \frac{2}{q_4^2} - \frac{{\bm q}_{23}\cdot{\bm q}_4}{q_{23}^2q_4^2} + \frac{3}{2}\frac{{\bm q}_{234}\cdot{\bm q}_{23}}{q_{234}^2q_{23}^2} + \frac{3}{2}\frac{{\bm q}_{234}\cdot{\bm q}_4}{q_{234}^2}\frac{{\bm q}_{23}\cdot{\bm q}_4}{q_{23}^2q_4^2} \bigg) \bigg( \frac{2}{q_2^2} - \frac{{\bm q}_2\cdot{\bm q}_3}{q_2^2q_3^2} + \frac{3}{2}\frac{{\bm q}_{23}\cdot{\bm q}_2}{q_{23}^2q_2^2} + \frac{3}{2}\frac{{\bm q}_{23}\cdot{\bm q}_3}{q_{23}^2}\frac{{\bm q}_2\cdot{\bm q}_3}{q_2^2q_3^2} \bigg)
\nonumber\\
& - \frac{25}{8} \frac{{\bm q}_1\cdot{\bm q}_{234}}{q_{34}^2q_{234}^2} \bigg( -2 + \frac{{\bm q}_2\cdot{\bm q}_{34}}{q_2^2} - \frac{3}{2}\frac{{\bm q}_{234}\cdot{\bm q}_2}{q_{234}^2}\frac{q_{34}^2}{q_2^2} - \frac{3}{2}\frac{{\bm q}_{234}\cdot{\bm q}_{34}}{q_{234}^2}\frac{{\bm q}_2\cdot{\bm q}_{34}}{q_{34}^2} \bigg) \frac{3{\bm q}_3\cdot{\bm q}_4 + 8q_4^2}{q_3^2q_4^2}
\nonumber\\
& + \frac{25}{4} \frac{{\bm q}_1\cdot{\bm q}_{234}}{q_{234}^2} \frac{1}{q_3^2q_4^2} \bigg[ -4 + 4\frac{{\bm q}_2\cdot{\bm q}_3}{q_2^2} - \frac{3}{2} \bigg( \frac{{\bm q}_{234}\cdot{\bm q}_2}{q_{234}^2}\frac{{\bm q}_3\cdot{\bm q}_4}{q_2^2} + 3\frac{{\bm q}_{234}\cdot{\bm q}_4}{q_{234}^2}\frac{{\bm q}_2\cdot{\bm q}_3}{q_2^2}
+ 4\frac{{\bm q}_{234}\cdot{\bm q}_3}{q_{234}^2}\frac{{\bm q}_2\cdot{\bm q}_3}{q_2^2} + 4\frac{{\bm q}_{234}\cdot{\bm q}_2}{q_{234}^2}\frac{q_3^2}{q_2^2} \bigg) \bigg]
\nonumber\\
& - \frac{25}{4} \frac{{\bm q}_1\cdot{\bm q}_2}{q_2^2q_{34}^2} \frac{3{\bm q}_3\cdot{\bm q}_4 + 8q_4^2}{q_3^2q_4^2} + \frac{25}{4}\frac{{\bm q}_1\cdot{\bm q}_{34}}{q_2^2q_{34}^2} \bigg[ -\frac{2}{q_4^2} + \frac{{\bm q}_3\cdot{\bm q}_4}{q_3^2q_4^2} - \frac{3}{2}\frac{{\bm q}_{34}\cdot{\bm q}_3}{q_{34}^2q_3^2} - \frac{3}{2}\frac{{\bm q}_{34}\cdot{\bm q}_4}{q_{34}^2}\frac{{\bm q}_3\cdot{\bm q}_4}{q_3^2q_4^2} \bigg] - \frac{25}{q_3^2q_4^2}\frac{{\bm q}_1\cdot{\bm q}_2}{q_2^2} \, ,
\\
& B_2 = \bigg( -\frac{2}{3} - \frac{{\bm q}_1\cdot{\bm q}_{234}}{q_1^2} + \frac{k^2}{q_{234}^2}\frac{{\bm q}_1\cdot{\bm q}_{234}}{q_1^2} \bigg)
\nonumber\\
& \hspace{0.3cm} \times \bigg\{ \frac{25}{4} \bigg( \frac{2}{q_4^2} - \frac{{\bm q}_{23}\cdot{\bm q}_4}{q_{23}^2q_4^2} + \frac{3}{2}\frac{{\bm q}_{234}\cdot{\bm q}_{23}}{q_{234}^2q_{23}^2} + \frac{3}{2}\frac{{\bm q}_{234}\cdot{\bm q}_4}{q_{234}^2}\frac{{\bm q}_{23}\cdot{\bm q}_4}{q_{23}^2q_4^2} \bigg) \bigg( \frac{2}{q_3^2} - \frac{{\bm q}_2\cdot{\bm q}_3}{q_2^2q_3^2} + \frac{3}{2}\frac{{\bm q}_{23}\cdot{\bm q}_2}{q_{23}^2q_2^2} + \frac{3}{2}\frac{{\bm q}_{23}\cdot{\bm q}_3}{q_{23}^2}\frac{{\bm q}_2\cdot{\bm q}_3}{q_2^2q_3^2} \bigg)
\nonumber\\
& \hspace{0.8cm} - \frac{25}{8q_{34}^2} \bigg( -2 + \frac{{\bm q}_2\cdot{\bm q}_{34}}{q_2^2} - \frac{3}{2}\frac{{\bm q}_{234}\cdot{\bm q}_2}{q_{234}^2}\frac{q_{34}^2}{q_2^2} - \frac{3}{2}\frac{{\bm q}_{234}\cdot{\bm q}_{34}}{q_{234}^2}\frac{{\bm q}_2\cdot{\bm q}_{34}}{q_{34}^2} \bigg) \frac{3{\bm q}_3\cdot{\bm q}_4 + 8q_4^2}{q_3^2q_4^2}
\nonumber\\
& \hspace{0.8cm} + \frac{25}{4q_3^2q_4^2} \bigg[ -4 + 4\frac{{\bm q}_2\cdot{\bm q}_3}{q_2^2} - \frac{3}{2} \bigg( \frac{{\bm q}_{234}\cdot{\bm q}_2}{q_{234}^2}\frac{{\bm q}_3\cdot{\bm q}_4}{q_2^2} + 3\frac{{\bm q}_{234}\cdot{\bm q}_4}{q_{234}^2}\frac{{\bm q}_2\cdot{\bm q}_3}{q_2^2} + 4\frac{{\bm q}_{234}\cdot{\bm q}_3}{q_{234}^2}\frac{{\bm q}_2\cdot{\bm q}_3}{q_2^2} + 4\frac{{\bm q}_{234}\cdot{\bm q}_2}{q_{234}^2}\frac{q_3^2}{q_2^2} \bigg) \bigg] \bigg\}
\nonumber\\
& - \frac{25}{8q_{34}^2} \bigg( -\frac{2}{3}\frac{{\bm q}_1\cdot{\bm q}_2}{q_1^2} + 4\frac{{\bm q}_1\cdot{\bm q}_2}{q_1^2}\frac{{\bm k}\cdot{\bm q}_2}{q_2^2} - \frac{4}{3}\frac{{\bm k}\cdot{\bm q}_2}{q_2^2} \bigg) \frac{3{\bm q}_3\cdot{\bm q}_4 + 8q_4^2}{q_3^2q_4^2}
\nonumber\\
& + \frac{25}{4q_1^2} \bigg( -\frac{2}{q_4^2} + \frac{{\bm q}_3\cdot{\bm q}_4}{q_3^2q_4^2} - \frac{3}{2}\frac{{\bm q}_{34}\cdot{\bm q}_3}{q_{34}^2q_3^2} - \frac{3}{2}\frac{{\bm q}_{34}\cdot{\bm q}_4}{q_{34}^2}\frac{{\bm q}_3\cdot{\bm q}_4}{q_3^2q_4^2} \bigg)
\nonumber\\
& \hspace{1cm} \times \bigg( 4\frac{{\bm k}\cdot{\bm q}_2}{q_2^2}\frac{{\bm q}_2\cdot{\bm q}_{34}}{q_{34}^2} - \frac{4}{3}\frac{{\bm k}\cdot{\bm q}_2}{q_2^2} + 4\frac{{\bm k}\cdot{\bm q}_{34}}{q_{34}^2}\frac{{\bm q}_2\cdot{\bm q}_{34}}{q_2^2} - \frac{4}{3}\frac{{\bm k}\cdot{\bm q}_{34}}{q_{34}^2} - \frac{2}{3}\frac{{\bm q}_2\cdot{\bm q}_{34}}{q_{34}^2} - \frac{8}{3}\frac{{\bm q}_2\cdot{\bm q}_{34}}{{q_2}^2} \bigg)
\nonumber\\
& + \frac{25}{4q_1^2q_2^2} \bigg( 12\frac{{\bm q}_3\cdot{\bm q}_4}{q_3^2}\frac{{\bm k}\cdot{\bm q}_4}{q_4^2} - 4\frac{{\bm k}\cdot{\bm q}_4}{q_4^2} - 4\frac{{\bm q}_3\cdot{\bm q}_4}{q_3^2} + 2\frac{{\bm q}_1\cdot{\bm q}_2}{q_3^2}\frac{{\bm q}_3\cdot{\bm q}_4}{q_4^2} + \frac{2}{3}\frac{{\bm q}_1\cdot{\bm q}_3}{q_3^2}\frac{{\bm q}_2\cdot{\bm q}_4}{q_4^2} \bigg)
\nonumber\\
& + \frac{25}{4} \bigg( \frac{2}{3} - \frac{{\bm k}\cdot{\bm q}_{34}}{q_{34}^2} + \frac{{\bm k}\cdot{\bm q}_{34}}{q_{34}^2}\frac{{\bm q}_{12}\cdot{\bm q}_{34}}{q_{12}^2} \bigg)
\nonumber\\
& \hspace{1cm} \times \bigg( \frac{2}{q_2^2} - \frac{{\bm q}_1\cdot{\bm q}_2}{q_1^2q_2^2} + \frac{3}{2}\frac{{\bm q}_{12}\cdot{\bm q}_1}{q_{12}^2q_1^2} + \frac{3}{2}\frac{{\bm q}_{12}\cdot{\bm q}_2}{q_{12}^2}\frac{{\bm q}_1\cdot{\bm q}_2}{q_1^2q_2^2} \bigg) \bigg( \frac{2}{q_4^2} - \frac{{\bm q}_3\cdot{\bm q}_4}{q_3^2q_4^2} + \frac{3}{2}\frac{{\bm q}_{34}\cdot{\bm q}_3}{q_{34}^2q_3^2} + \frac{3}{2}\frac{{\bm q}_{34}\cdot{\bm q}_4}{q_{34}^2}\frac{{\bm q}_3\cdot{\bm q}_4}{q_3^2q_4^2} \bigg) \, .
\end{align}
The fully symmetric kernels can be obtained by taking into account possible permutations,
\begin{align}
F_4^{(s)}({\bm q}_1,{\bm q}_2,{\bm q}_3,{\bm q}_4) = & \frac{1}{4!} \Big[ F_4({\bm q}_1,{\bm q}_2,{\bm q}_3,{\bm q}_4) + \text{(23 cyclic)} \Big]
\, ,
\\
G_4^{(s)}({\bm q}_1,{\bm q}_2,{\bm q}_3,{\bm q}_4) = & \frac{1}{4!} \Big[ G_4({\bm q}_1,{\bm q}_2,{\bm q}_3,{\bm q}_4) + \text{(23 cyclic)} \Big] \, .
\end{align}
Note that the equations and solutions up to third order can be found in~\cite{Jeong:2010ag}.


\begin{thebibliography}{99}


\bibitem{wmap}
  C.~L.~Bennett {\it et al.}  [WMAP Collaboration],
  %``Nine-Year Wilkinson Microwave Anisotropy Probe (WMAP) Observations: Final Maps and Results,''
  Astrophys.\ J.\ Suppl.\  {\bf 208}, 20 (2013)
  [arXiv:1212.5225 [astro-ph.CO]]~;
  %%CITATION = ARXIV:1212.5225;%%
  G.~Hinshaw {\it et al.}  [WMAP Collaboration],
  %``Nine-Year Wilkinson Microwave Anisotropy Probe (WMAP) Observations: Cosmological Parameter Results,''
  Astrophys.\ J.\ Suppl.\  {\bf 208}, 19 (2013)
  [arXiv:1212.5226 [astro-ph.CO]].
  %%CITATION = ARXIV:1212.5226;%%


\bibitem{planck}
  P.~A.~R.~Ade {\it et al.}  [Planck Collaboration],
  %``Planck 2013 results. I. Overview of products and scientific results,''
  arXiv:1303.5062 [astro-ph.CO]~;
  %%CITATION = ARXIV:1303.5062;%%
  P.~A.~R.~Ade {\it et al.}  [Planck Collaboration],
  %``Planck 2013 results. XVI. Cosmological parameters,''
  arXiv:1303.5076 [astro-ph.CO].
  %%CITATION = ARXIV:1303.5076;%%


\bibitem{PIXIE}
  A.~Kogut, D.~J.~Fixsen, D.~T.~Chuss, J.~Dotson, E.~Dwek, M.~Halpern, G.~F.~Hinshaw and S.~M.~Meyer {\it et al.},
  %``The Primordial Inflation Explorer (PIXIE): A Nulling Polarimeter for Cosmic Microwave Background Observations,''
  JCAP {\bf 1107}, 025 (2011)
  [arXiv:1105.2044 [astro-ph.CO]].
  %%CITATION = ARXIV:1105.2044;%%
  %62 citations counted in INSPIRE as of 27 Feb 2014


\bibitem{PRISM}
  P.~André {\it et al.}  [ PRISM Collaboration],
  %``PRISM (Polarized Radiation Imaging and Spectroscopy Mission): An Extended White Paper,''
  arXiv:1310.1554 [astro-ph.CO].
  %%CITATION = ARXIV:1310.1554;%%
  %7 citations counted in INSPIRE as of 27 Feb 2014


\bibitem{LiteBIRD}
  T.~Matsumura, Y.~Akiba, J.~Borrill, Y.~Chinone, M.~Dobbs, H.~Fuke, A.~Ghribi and M.~Hasegawa {\it et al.},
  %``Mission design of LiteBIRD,''
  arXiv:1311.2847 [astro-ph.IM].
  %%CITATION = ARXIV:1311.2847;%%
  %1 citations counted in INSPIRE as of 27 Feb 2014


\bibitem{Anderson:2013zyy}
  L.~Anderson {\it et al.}  [BOSS Collaboration],
  %``The clustering of galaxies in the SDSS-III Baryon Oscillation Spectroscopic Survey: Baryon Acoustic Oscillations in the Data Release 10 and 11 galaxy samples,''
  arXiv:1312.4877 [astro-ph.CO].
  %%CITATION = ARXIV:1312.4877;%%
  %9 citations counted in INSPIRE as of 27 Feb 2014


\bibitem{Blake:2011en}
  C.~Blake, E.~Kazin, F.~Beutler, T.~Davis, D.~Parkinson, S.~Brough, M.~Colless and C.~Contreras {\it et al.},
  %``The WiggleZ Dark Energy Survey: mapping the distance-redshift relation with baryon acoustic oscillations,''
  Mon.\ Not.\ Roy.\ Astron.\ Soc.\  {\bf 418}, 1707 (2011)
  [arXiv:1108.2635 [astro-ph.CO]].
  %%CITATION = ARXIV:1108.2635;%%
  %241 citations counted in INSPIRE as of 27 Feb 2014


\bibitem{delaTorre:2013rpa}
  S.~de la Torre, L.~Guzzo, J.~A.~Peacock, E.~Branchini, A.~Iovino, B.~R.~Granett, U.~Abbas and C.~Adami {\it et al.},
  %``The VIMOS Public Extragalactic Redshift Survey (VIPERS). Galaxy clustering and redshift-space distortions at z=0.8 in the first data release,''
  arXiv:1303.2622 [astro-ph.CO].
  %%CITATION = ARXIV:1303.2622;%%
  %23 citations counted in INSPIRE as of 27 Feb 2014


\bibitem{MSDESI}
http://desi.lbl.gov


\bibitem{HETDEX}
http://www.hetdex.org


\bibitem{LSST}
http://www.lsst.org/lsst/


\bibitem{Euclid}
http://sci.esa.int/euclid/


\bibitem{nGreviews}
For a recent collection of reviews, see e.g.
Class.\ Quant.\ Grav.\  {\bf 27}, ``Focus section on non-linear and non-Gaussian cosmological perturbations'' (2010)~;
Adv.\ Astron.\ {\bf 2010}, ``Testing the Gaussianity and Statistical Isotropy of the Universe'' (2010).


\bibitem{nlpowerspectrum}
%\cite{Jain:1993jh}
%\bibitem{Jain:1993jh}
  B.~Jain and E.~Bertschinger,
  %``Second order power spectrum and nonlinear evolution at high redshift,''
  Astrophys.\ J.\  {\bf 431}, 495 (1994)
  [astro-ph/9311070]~;
  %%CITATION = ASTRO-PH/9311070;%%
%\cite{Jeong:2006xd}
%\bibitem{Jeong:2006xd}
  D.~Jeong and E.~Komatsu,
  %``Perturbation Theory Reloaded: Analytical Calculation of Non-linearity in
  %Baryonic Oscillations in the Real Space Matter Power Spectrum,''
  Astrophys.\ J.\  {\bf 651}, 619 (2006)
  [arXiv:astro-ph/0604075].
  %%CITATION = ASJOA,651,619;%%


%\cite{Jeong:2010ag}
\bibitem{Jeong:2010ag}
  D.~Jeong, J.~O.~Gong, H.~Noh and J.~c.~Hwang,
  %``General relativistic effects on non-linear power spectra,''
  Astrophys.\ J.\  {\bf 727}, 22 (2011)
  [arXiv:1010.3489 [astro-ph.CO]].
  %%CITATION = ASJOA,727,22;%%


%\cite{Bernardeau:2001qr}
\bibitem{Bernardeau:2001qr}
  F.~Bernardeau, S.~Colombi, E.~Gaztanaga and R.~Scoccimarro,
  %``Large-scale structure of the universe and cosmological perturbation
  %theory,''
  Phys.\ Rept.\  {\bf 367}, 1 (2002)
  [arXiv:astro-ph/0112551].
  %%CITATION = PRPLC,367,1;%%


\bibitem{Weinberg:2012es}
  D.~H.~Weinberg, M.~J.~Mortonson, D.~J.~Eisenstein, C.~Hirata, A.~G.~Riess and E.~Rozo,
  %``Observational Probes of Cosmic Acceleration,''
  Phys.\ Rept.\  {\bf 530}, 87 (2013)
  [arXiv:1201.2434 [astro-ph.CO]].
  %%CITATION = ARXIV:1201.2434;%%
  %138 citations counted in INSPIRE as of 28 Feb 2014'


%\cite{Hwang:2012bi}
\bibitem{Hwang:2012bi}
  J.~-c.~Hwang, H.~Noh and J.~-O.~Gong,
  %``Second order solutions of cosmological perturbation in the matter dominated era,''
  Astrophys.\ J.\  {\bf 752}, 50 (2012)
  [arXiv:1204.3345 [astro-ph.CO]].
  %%CITATION = ARXIV:1204.3345;%%


\bibitem{gaugedep}
%\cite{Yoo:2009au}
%\bibitem{Yoo:2009au}
  J.~Yoo, A.~L.~Fitzpatrick and M.~Zaldarriaga,
  %``A New Perspective on Galaxy Clustering as a Cosmological Probe: General
  %Relativistic Effects,''
  Phys.\ Rev.\  D {\bf 80}, 083514 (2009)
  [arXiv:0907.0707 [astro-ph.CO]]~;
  %%CITATION = PHRVA,D80,083514;%%
%\cite{Bonvin:2011bg}
%\bibitem{Bonvin:2011bg}
  C.~Bonvin and R.~Durrer,
  %``What galaxy surveys really measure,''
  Phys.\ Rev.\ D {\bf 84}, 063505 (2011)
  [arXiv:1105.5280 [astro-ph.CO]]~;
  %%CITATION = ARXIV:1105.5280;%%
%\cite{Challinor:2011bk}
%\bibitem{Challinor:2011bk}
  A.~Challinor and A.~Lewis,
  %``The linear power spectrum of observed source number counts,''
  Phys.\ Rev.\ D {\bf 84}, 043516 (2011)
  [arXiv:1105.5292 [astro-ph.CO]]~;
  %%CITATION = ARXIV:1105.5292;%%
%\cite{Jeong:2011as}
%\bibitem{Jeong:2011as}
  D.~Jeong, F.~Schmidt and C.~M.~Hirata,
  %``Large-scale clustering of galaxies in general relativity,''
  Phys.\ Rev.\ D {\bf 85}, 023504 (2012)
  [arXiv:1107.5427 [astro-ph.CO]].
  %%CITATION = ARXIV:1107.5427;%%


%\cite{Noh:2004bc}
\bibitem{Noh:2004bc}
  H.~Noh and J.~c.~Hwang,
  %``Second-order perturbations of the Friedmann world model,''
  Phys.\ Rev.\  D {\bf 69}, 104011 (2004)
  [arXiv:astro-ph/0305123].
  %%CITATION = PHRVA,D69,104011;%%


%\cite{Arnowitt:1962hi}
\bibitem{Arnowitt:1962hi}
  R.~L.~Arnowitt, S.~Deser and C.~W.~Misner,
  %``The dynamics of general relativity,''
  arXiv:gr-qc/0405109.
  %%CITATION = GR-QC/0405109;%%


%\cite{Bardeen:1980kt}
\bibitem{Bardeen:1980kt}
  J.~M.~Bardeen,
  %``Gauge Invariant Cosmological Perturbations,''
  Phys.\ Rev.\  D {\bf 22}, 1882 (1980).
  %%CITATION = PHRVA,D22,1882;%%



\bibitem{Scoccimarro:1996jy}
  R.~Scoccimarro,
  %``Cosmological perturbations: Entering the nonlinear regime,''
  Astrophys.\ J.\  {\bf 487}, 1 (1997)
  [astro-ph/9612207].
  %%CITATION = ASTRO-PH/9612207;%%


%\cite{Komatsu:2010fb}
\bibitem{Komatsu:2010fb}
  E.~Komatsu {\it et al.}  [WMAP Collaboration],
  %``Seven-Year Wilkinson Microwave Anisotropy Probe (WMAP) Observations:
  %Cosmological Interpretation,''
  Astrophys.\ J.\ Suppl.\  {\bf 192}, 18 (2011)
  [arXiv:1001.4538 [astro-ph.CO]].
  %%CITATION = APJSA,192,18;%%


\end{thebibliography}
\end{document}